\documentclass[]{mn2e}

\usepackage[caption=false]{subfig}
\usepackage{graphicx}
\usepackage{amsmath}
\usepackage{amssymb}

\def\kms{km~s$^{-1}$ }
\def\etal{et al.\,}

\def\eg{e.g., \,}
\def\R200{$R_{200}$}
\def\MSYR{M$_\odot$yr$^{-1}$}

\title[The Double Galaxy Cluster Abell 2465 II.]
{The Double Galaxy Cluster
Abell 2465 II. Star Formation in the Cluster}

\author[G.  A. Wegner, D. S. Chu, H. S. Hwang]
{ Gary A. Wegner$^1$ Devin S. Chu$^1$, Ho Seong Hwang$^2$\\
$^1$Department of Physics \& Astronomy, Dartmouth College, 6127 Wilder
Laboratory, Hanover, NH 03755, U.S.A.\\
$^2$School of Physics, Korea Institute for Advanced Study, 85 Hoegiro, Dongdaemun-go,
Seoul 130-722, Republic of Korea
}

\date{Accepted ; in original form }

\pagerange{\pageref{firstpage}--\pageref{lastpage}} \pubyear{2010}
\begin{document}
\label{firstpage}
\maketitle

\begin{abstract}
We investigate the star formation rate and its
location in the major merger cluster
Abell 2465 at $z$ = 0.245. Optical properties of the cluster are
described in Paper I.
Measurements of the H$\alpha$ and infrared dust emission of galaxies in the cluster were made 
with an interference filter centred on the redshifted line 
at a wavelength of 817 nm and utilized data from the WISE satellite 12 $\mu$m band. Imaging 
in the Johnson $U$ and $B$ bands was obtained, and along with SDSS $u$ and $r$
was used to study the blue fraction, which appears enhanced, as a further signatures of 
star formation in the cluster. Star formation rates were calculated using standard 
calibrations. The total star formation rate normalized by the cluster mass,
$\Sigma SFR/M_{cl}$ compared to compilations for other clusters indicate 
that the components of Abell 2465 lie above the mean $z$ and
$M_{cl}$ relations, suggestive that interacting galaxy clusters have enhanced star formation.
The projected radial distribution of the star forming galaxies does not follow a NFW 
profile and is relatively flat indicating that fewer star forming galaxies are in the cluster
centre. The morphologies of the H$\alpha$ sources within $R_{200}$
for the cluster as a whole indicate that many are disturbed or
merging, suggesting that a combination of merging or harassment is
working.  
\end{abstract}

\begin{keywords}{galaxies: clusters: general -- galaxies:
clusters: individual: Abell 2465}; galaxies: evolution; galaxies: starburst
\end{keywords}

\section{INTRODUCTION}
Merging galaxy clusters can provide information on several aspects of cosmology. 
One sees the interaction of different matter components
in their gravitational fields and is provided clues about galaxy evolution.
The dynamics in galaxy
cluster collisions including the galaxies', baryonic and dark matter components,
have been studied by several authors (\eg Roettiger \etal 1996, 1997; Ricker 1998; 
Tazikawa 2000;
Ricker \& Sarazin 2001; Ritchie \& Thomas 2002;
Springel \& Farrar 2007; Mastropietro
\& Burkert 2008; Poole \etal 2008; Planelles \& Quilis 2009; Vijayaraghavan \& Ricker 2013) 
although the effects
of cluster collisions on star formation rates (SFR) have received less attention. 

It is generally held that galaxy clusters grow from merging
and their less relaxed exterior regions are more gas rich whilst interior sections having
undergone more interactions are gas depleted, leading to differences 
in star formation.
For most low redshift ($z~\la~  1)$ galaxy clusters in the local universe
the total SFR per cluster mass drops with decreasing $z$, 
$\Sigma SFR/M_{cl} \sim (1 + z)^6$ 
and also cluster mass, $\sim M_{cl}^{-1.5}$
(\eg Finn \etal 2005; Koyama \etal 2010; Chung \etal 2011; 
Popesso \etal 2012; Webb \etal 2013). This is related to the general process 
of 'downsizing' in which the star formation decreases
with the lowering of redshift (\eg Le Floc'h \etal 2005) and the SFR-surface density
relation (\eg Martinex \etal 2002; G\'{o}mez \etal 2003). Using X-ray and optical data, 
Lagana \etal (2008) found
that massive 'hot' clusters have less star formation than 'cold' clusters of lower mass.

However in collisions between clusters, enhancements of star formation 
have been reported at lower redshifts (\eg Miller 2005; Cortese \etal 2004; 
Ferrari \etal 2005;
Miller \& Owen 2003; Hwang \& Lee 2009; Chung \etal 2009; 2010; Rawle \etal 2010;
Ma \etal 2010; Russell \etal 2010; Biviano \etal 2011; Bourdin \etal 2011; Owers \etal 2011;
Canning \etal 2012; Dawson \etal 2012; Rawle \etal 2014). 
Not all objects show this effect, however, and it is of interest to compare the 
regions of star formation and the distributions of the
displaced baryonic and dark matter components produced by the cluster interactions.

Mechanisms contributing to the star formation 
in galaxy evolution are complicated, the dominant one for a particular object
probably depends on the environment, and include ram pressure stripping,
strangulation, galaxy harassment, galaxy-galaxy 
interactions, and tidal effects on the cluster galaxies (Bekki 1999, 2013; 
Gnedin 2003a,b).
Martig \& Bournaud (2008) found that star formation in mergers can be enhanced by
the tidal field in galaxy clusters and complicated
substructure can induce appreciable effects. 

Substructure in galaxy clasters has been long known (\eg Geller \& Beers 1982) and used to
study their formation and relaxation (\eg Burgett \etal 2004). Investigations of the star
formation rates among galaxies in the substructure of clusters (\eg Cohen \etal 2014; Cohen,
Hickox, \& Wegner 2014) are complicated due to the difficulties of assessing their dynamical
states, but show a weak but significant correlation with substructure in the sense that the
unrelaxed clusters, as defined by the number of their subcomponents, have higher star 
formation which also increases outwards with clustercentric distance.

Interpretation is further complicated by a lack of knowledge of the phase and geometry 
of the collision when the clusters are observed, \eg 
before or after closest core passage, whether it is a first or second passage
of the merger, or if the orbit is linear or parabolic. 
Bekki (2013) finds that ram pressure stripping enhances or 
reduces star formation in groups and clusters of galaxies depending on their 
initial morphologies and geometrical 
conditions. Vijayaraghavan \& Ricker (2013) have modelled ram pressure stripping in 
cluster collisions and find that it can affect galaxies at large distances as well as
stopping star formation in most central regions of clusters.
 

This paper examines star formation in the double galaxy cluster Abell 2465.
It has the relatively simple structure
of two subclusters, referred to here as the SW and NE clumps.
It is a double X-ray source and the basic properties are given
in Wegner (2011; hereafter Paper I) 
where $R_{200} \approx 1.2$ Mpc and $M_{200} \approx 4 \times 10^{14} M_\odot$
for each subcluster although their other properties differ.
In Paper I an apparent overabundance of galaxies showing H$\alpha$ emission
was suggested; 37 percent of the
158 spectroscopically observed cluster members show emission lines and
fall predominantly in the star forming region of the 
$\log([NII]/H \alpha), \log([OIII]/H \beta)$ diagnostic diagram. 
Most of these sources are closer to
the cluster centres, whereas in single galaxy clusters, emission line galaxies 
tend to be more numerous in the outer infall regions of the clusters (\eg 
Balogh \etal 2004; Rines \etal 2005).

The detection of H$\alpha$ emission
photometrically by combining 
narrow and broad band filters is employed, a technique used by many authors, including
Kennicut \& Kent (1983),
Salzer \etal (2005), Werk \etal (2010), Kellar \etal (2012). In particular,
Finn \etal (2005), Morioka \etal (2008), Shioya \etal (2008), Westra \& Jones (2008),
James \etal (2004), Fujita \etal (2003), Kodama \etal (2004), and 
Koyama \etal (2010, 2011) have targeted galaxy clusters at intermediate
redshifts employing this technique. For Abell 2465, 
the redshift is $z = 0.2453$ and H$\alpha$ appears at a wavelength near 817 nm with individual
cluster members emitting between 814 and 820 nm. 

The other well known technique for studying star formation is using infrared dust emission.
Abell 2465 is covered by the Wide-fileld Infrared Survey Explorer (WISE; Wright \etal 2010) 
satellite and although the galaxy cluster is near the 
sensitivity limit of this instrument, it provides a valuable check on the H$\alpha$ 
measurement.

In addition to $r', i'$, and $g'$ imaging obtained by the CFHT and
described in Paper I and Wegner \etal (in preparation) Abell 2465 is now
included in the Sloan Digital Sky Survey data release 9 (DR9; Ahn \etal 2012).
For this study, $U$ and $B$ observations were
secured at the MDM Observatory to further study possible star forming
galaxies. These data can be used for studying the blue fraction of the galaxies in 
the cluster which can also be taken as an indicator of the SFR.

This paper reports H$\alpha$ and $UB$ imaging in Abell 2465,
used to estimate the extent of star formation in this double system. 
Section 2 describes the H$\alpha$ observations. Section 3 gives the resulting H$\alpha$ SFR.
Section 4 gives the SFR determined from WISE data and Section 5 compares these
results with star formation rates in other galaxy clusters. Section 6
discusses the morphologies of the galaxies. Section 7 describes the $U$ and $B$
measurements and colour-colour and colour-magnitude diagrams qnd 
blue fraction. Section 8 discusses
possible mechanisms for causing star formation and Section 9 gives conclusions.  
The WMAP 5-year cosmological parameters as implemented in NED are used in this paper.

\section{The H$\alpha$ imaging}
\begin{figure}
\includegraphics[width=8cm]{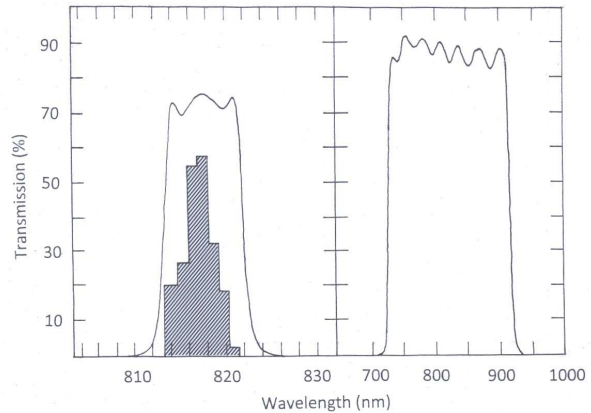}
\begin{footnotesize}
\caption{Transmissions of the filters used for the H$\alpha$
measurement. The narrow band filter centred on 817 nm, the wavelength of
H$\alpha$ at redshift $z = 0.245$ for Abell 2465 (left) and the broad
band $i_g$ filter (right). Both transmissions supplied by the manufacturers.
The shaded histogram in the left panel represents the spectroscopically
measured redshifts in Wegner(2011).
}
\end{footnotesize}
\label{filters}
\end{figure}

The wavelength of the H$\alpha$ line at the redshift of Abell 2465 is near 817 nm 
in a clear spectral region between the many telluric emission lines. A
custom narrow band filter for observing H$\alpha$ was obtained from  
the Andover Corp. It has a peak transmission at 817.7 nm 
($m_{817}$) and a 
full width half maximum (FWHM) of 8.77 nm. The wide filter was a Gunn $i$
($i_g$) filter with nearly the same central wavelength of 820 nm and a FWHM of
185 nm, was manufactured by Custom Scientific. The transmissions of the two filters are
given in Figure~\ref{filters} and compared with the spectroscopic 
redshifts measured in the cluster.

\subsection{The H$\alpha$ observations and reductions}
H$\alpha$ observations of Abell 2465 were obtained 2012 September 19-23  using the 
2.4 m Hiltner telescope at the MDM Observatory on Kitt Peak. The `Nellie'
CCD was used\footnote{Further details of these instruments can be found on the MDM
Observatory's web-page: http://www.astro.lsa.umich.edu/obs/mdm/technical/index.html.
}. This is a thick front illuminated Tektronix 2048 $\times$ 2048 Space Telescope
Imaging Spectrograph (STIS) detector. The readout noise is 
4.38 e$^-$, the gain is
2.94 ADU, and with the 2.4 m telescope the image scale is 0.24 arcsec per pixel. 
With the 817 nm filter, the field is about 8.3 arcmin in 
diameter and the $i_g$ field is a square of the same size. 
A series of three 20
minute exposures were made of 18 overlapping fields in Abell 2465
with the 817 nm filter plus
two 5 minute exposures with the $i_g$ filter in the sequence
$m_{817} - i_g -m_{817} - i_g -m_{817}$ and the images were dithered by a few pixels.
Standard data reduction procedures employed IRAF \footnote{{\sc IRAF} is distributed by 
the National Optical Astronomy Observatories which are operated 
by the Association of Universities for
Research in Astronomy, Inc. under cooperative agreement with the National Science 
Foundation.
}. Ten bias frames
were taken every night, averaged, and subtracted from each image. Flat fields
were made inside the telescope dome. Four bad columns were corrected using
{\sc fixpix}. Each set of three $m_{817}$ images was averaged using {\sc imcombine} 
with positional offsets measured from matching stars on each frame and 
the {\sc ccdclip} option was employed for 
cosmic ray rejection. For the $i_g$ pairs,
cosmic rays were eliminated with the
{\sc lacosmic} program (van Dokkum 2001) before 
summing. The seeing FWHM measured on the images was $1.5~\pm ~0.1$ 
arcsec. 

World Coordinate System (WCS) coordinates were found using the IRAF programs
{\sc ccmap} for a first approximation followed with {\sc msctpeak} with the 
US Naval Observatory (USNO) b
catalog. This gave fits accurate to about $\pm 0.2$~arcsec for all averaged
images. 

The magnitudes of all objects on each frame were determined with the
program {\sc Sextractor} (Bertin \& Arnouts 1996) using the {\sc mag\_auto}
measurements. The $m_{817}$ image was first measured with the single image mode
and sources were matched by measuring the $i_g$ image with the double
image mode. The zeropoints were found from 
five refererence
stars on each image employing $i'$ band photometry from the SDSS DR9
which gave a mean zeropoint error of $\pm 0.03$
mag. per image for the comparison stars with $i' \approx 15 -16$~mag. Stars were
separated from galaxies using the {\sc Sextractor} stellarity parameters and objects
were further inspected by eye.

\subsection{Calibration of the H$\alpha$ emission}
The wide and narrow band filters provide two equations for
the continuum level, $C$, and the emission near H$\alpha$, $E$.
This is simplified by assuming rectangular responses and constant $C$.
The ratio of the FWHM of the two filters is $w = 185/8.77 = 21.09$,
which gives:
\begin{equation}F_{i_g} = \left(E + wC\right)~ \rm{and}~ 
F_{817} = \left(E + C\right) \end{equation}
where the zeropoints, $z_p$ were obtained
from the magnitudes using
$F_p = z_p10^{-0.4m_p}$. 

The galaxy 
J339.80887-5.74450 which has strong H$\alpha$ +
[N II] emission (EW = 82 {\AA} for H$\alpha$) and its
corresponding galaxy magnitudes are:
$$m_{817} = 18.567~ \rm{and}~ m_{i_g} = 19.215$$
Employing Anglo-Australian Telescope (AAT) and MDM data 
in Paper I, two cluster galaxies J339.85236-5.78806 and 
J339.91864-5.72389 calibrated the AAT data which were in common with calibrated MDM
spectra obtained 2008 September 26 and 27 on photometric nights 
with the {\it irscal} flux
standards Hiltner 102 and HD19445 included in the IRAF
{\sc onedspec} tasks {\sc standard}, {\sc sensfunc}, and {\sc calibrate}. 
Using the mean Kitt Peak
atmospheric extinction, the flux relative to J339.80887-5.74450 through 
the two filters is:
$$F_{817} = 1.876 \times 10^{-14} \times 10^{-0.4(m_{817}-18.567)}~\rm{and}$$
$$F_{i_g} = 1.818 \times 10^{-13} \times 10^{-0.4(m_{i_g}-19.215)}$$
{erg~s$^{-1}$ cm$^{-2}$
with estimated errors of $\pm 18$\%. 
 The continuum for each galaxy is 
obtained from:
$$C = \left( F_{i_g} - F_{817}\right)/\left(w-1\right)$$ 
which gives the emission component:
$$E = F_{817} - C$$
To correct for the 
[N II] emission at $\lambda \lambda$ 6548 and 6583, the 
final H$\alpha$ apparent luminosity is taken to be 
\begin{equation} F_{H\alpha} = 0.71 E \end{equation}
found using spectroscopic measurements of 42 galaxies in Paper I, comparable to
0.75 found by Kennicut(1983) and Tresse \etal (1999; 2002).

\subsection{Galaxy selection}
The same photometric standards were used for the $i_g$ and $m_{817}$ zero points, so the 
ratio $F_{817}/F_{i_g} \approx 1/w$ for a galaxy with no H$\alpha$ emission.
When H$\alpha$ is present, this ratio increases.
Plotting $F_{817}$ against $F_{i_g}$ produces the linear relation
in Figure~\ref{Fcuts}. A linear fit to the dense region
in Figure~\ref{Fcuts}a gives
a slope of $m = 1.00 \pm 0.01$ and an intercept $b = -1.25 \pm 0.05$ 
which checks the approximations above, which predict $b = -1.32$
from $\log w$. 

\begin{figure*}
\centering
\subfloat{\includegraphics[width=0.45\textwidth]{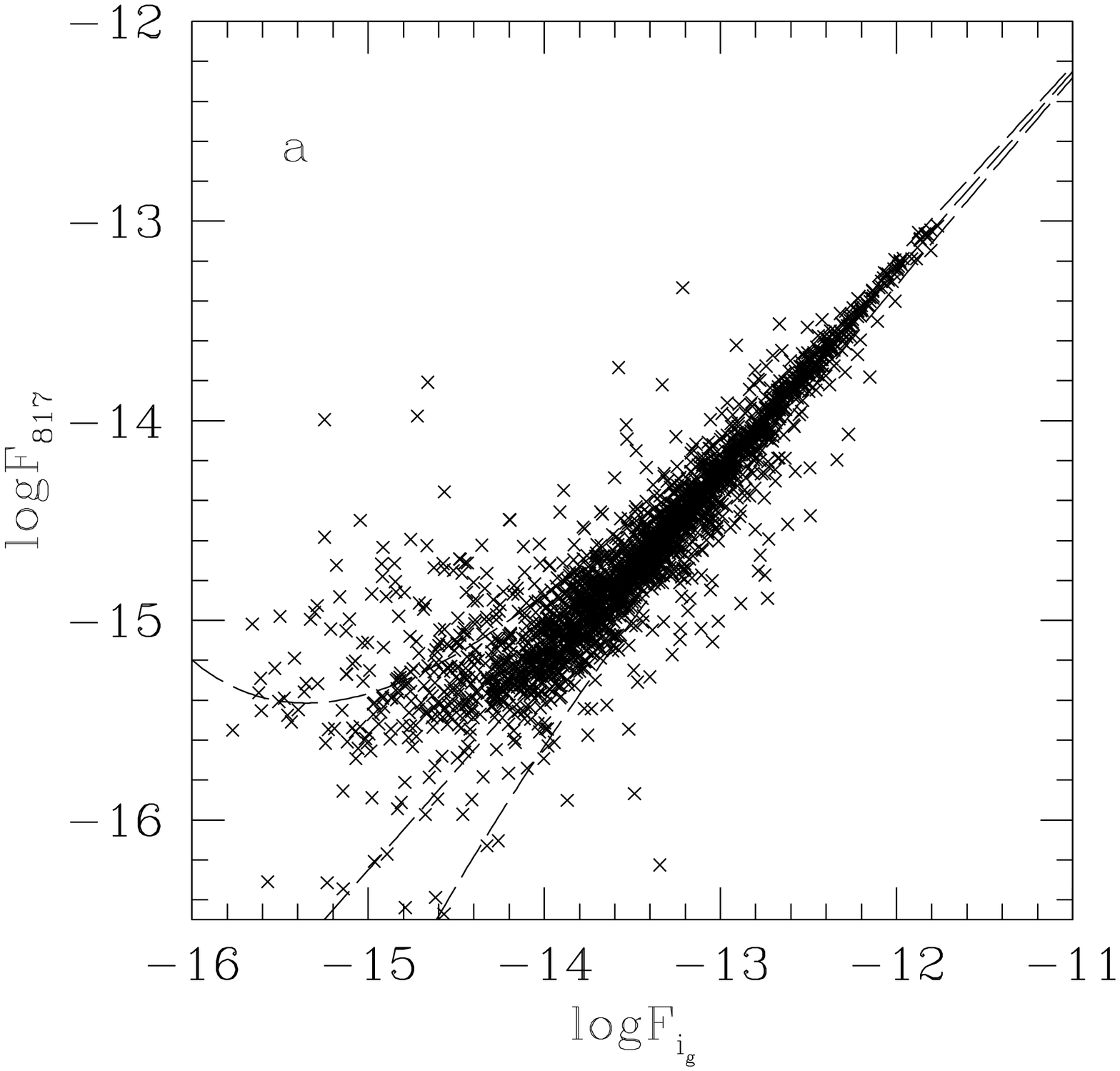}}
\subfloat{\includegraphics[width=0.45\textwidth]{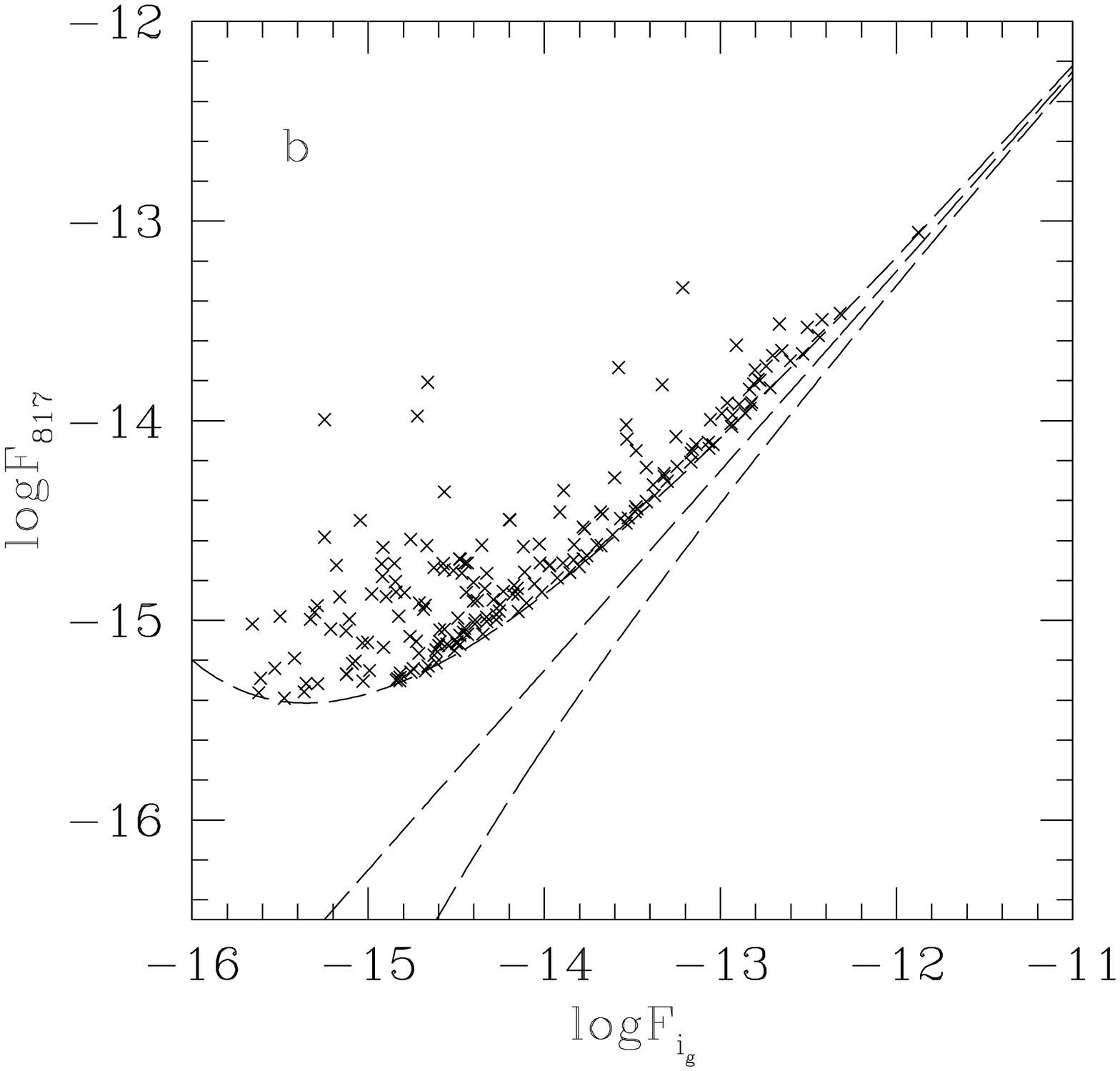}}

\subfloat{\includegraphics[width=0.45\textwidth]{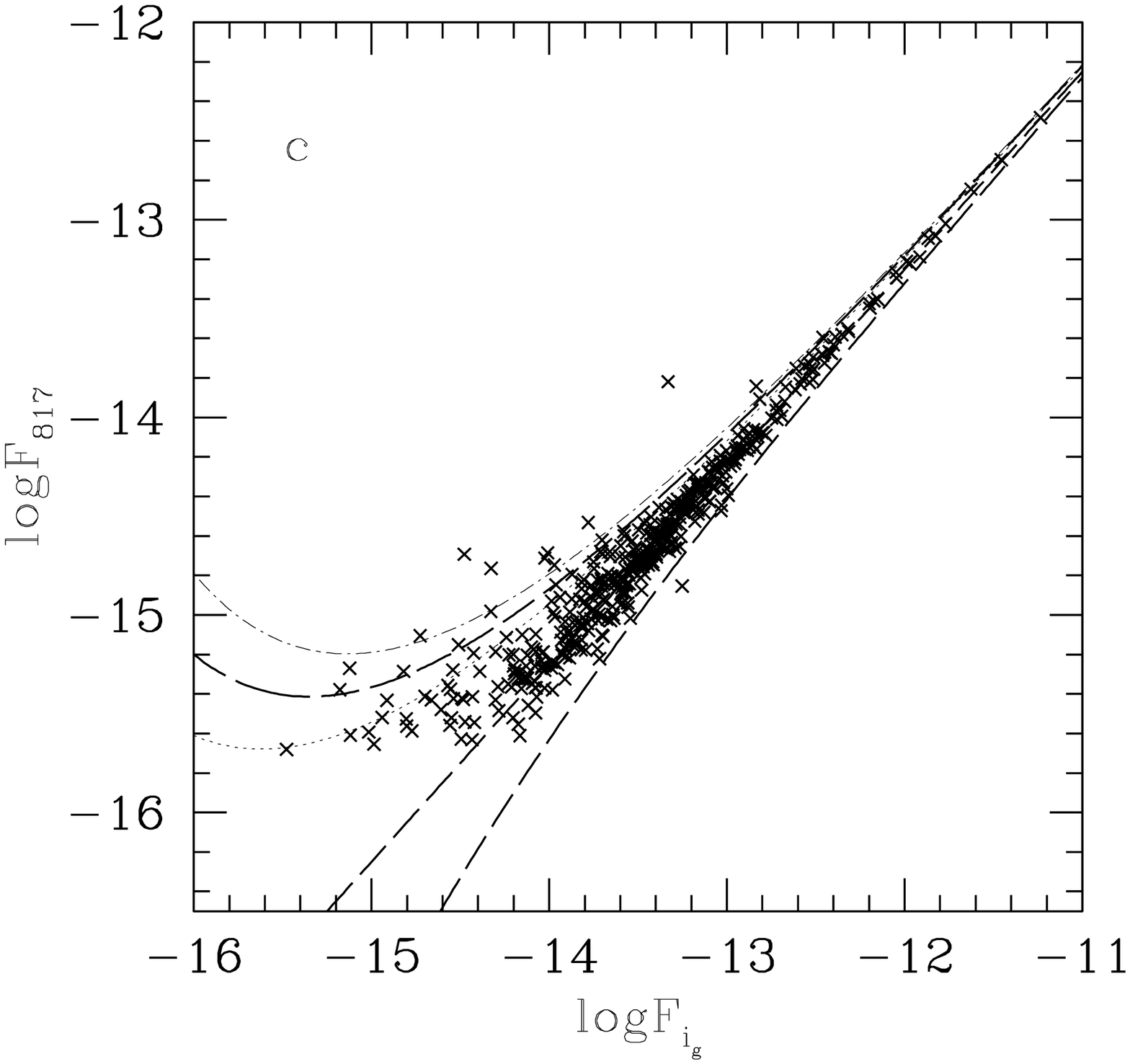}}
\subfloat{\includegraphics[width=0.45\textwidth]{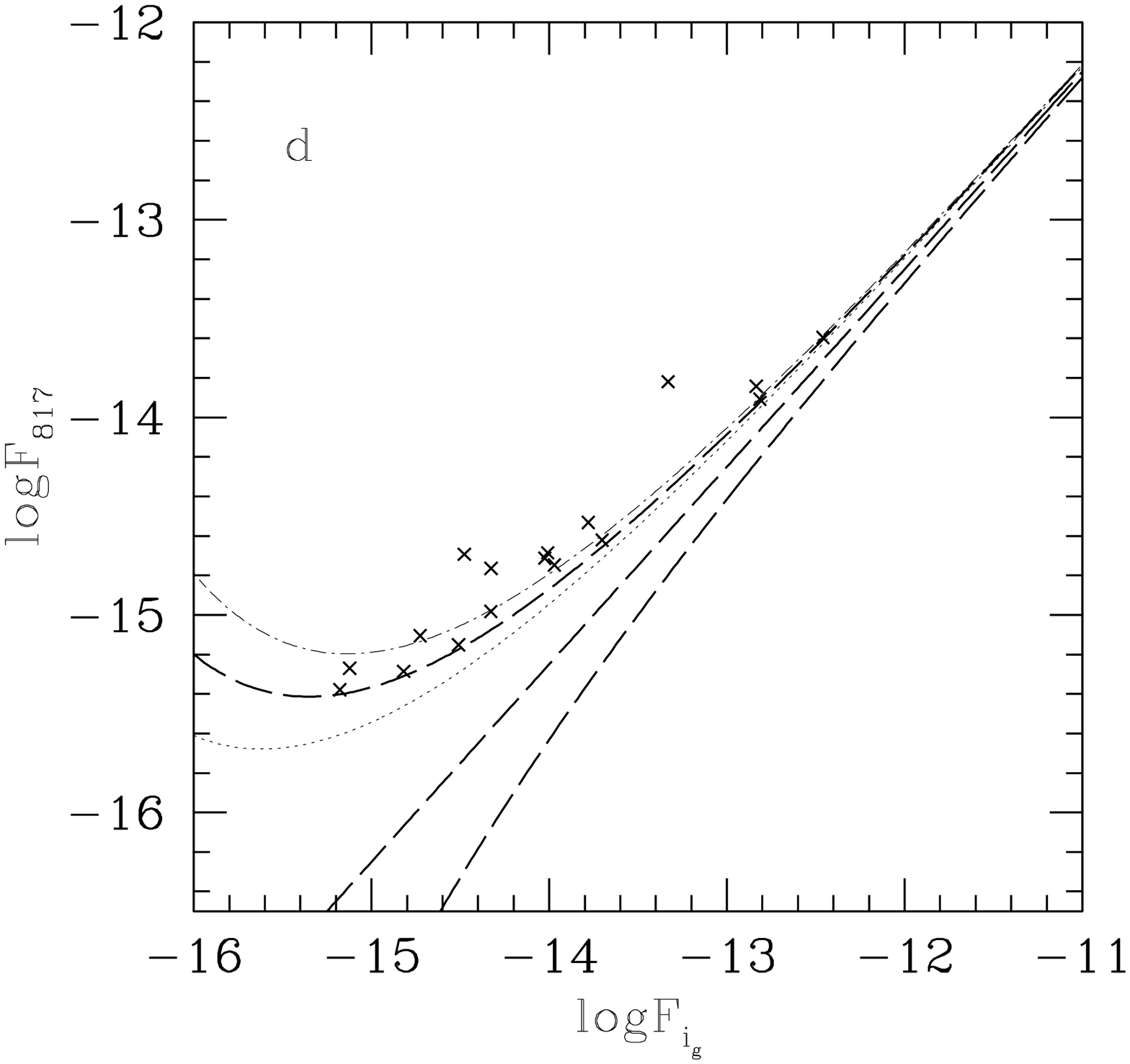}}

\caption{Showing the selection of the H$\alpha$ emission 
galaxies.The dashed curves show the locations of the fit to the 
non-emission galaxies in the centre and the estimated $\pm 2.5\mathfrak{c}$ curves. 
(a) all galaxies and (b) emission line galaxies selected above the $2.5\mathfrak{c}$ cut.
Bottom panels are for all galaxies in the background region (c) and 
after applying the same selection cut (d).The 2$\mathfrak{c}$ and 3$\mathfrak{c}$
cuts are indicated by the dotted curves in the two lower panels. The lower flux limit 
$\log F_{817} \approx -15.5$ corresponds to a magnitude $m_{817} \sim +23.0.$ 
}
\label{Fcuts}
\end{figure*}

The scatter in Figure~\ref{Fcuts} grows for fainter galaxies, but
sources with H$\alpha$ emission also produce asymmetry in
the vertical scatter seen in Figure~\ref{EWhisto} which is a histogram of all the
measured H$\alpha$ equivalent widths.
This was used to disentangle
the effects of the emission lines from the observational errors.
Above the line, the scatter is larger than below as 
emission line sources only occur above the line.
Absorption line galaxies below the line are weaker with smaller
equivalent widths, as demonstrated \eg by G{\'o}mez \etal (2003),
so the residuals below the line were used as an approximation
for the observational errors and fitted to Gaussians with maxima on the line.  
Their standard deviations are denoted here by $\mathfrak{c}$.
A function of $\log F_{i_g}$, 
$\mathfrak{c}$ was fit by
$\log \mathfrak{c} = -0.365 \log F_{i_g} - 5.93 $. In Figure~\ref{Fcuts}, the loci of
$\pm 2.5 \mathfrak{c}$ in the photometric error are shown as the dashed curves
and sources above this curve were selected as emission line sources.

\begin{figure}
\includegraphics[width=8cm]{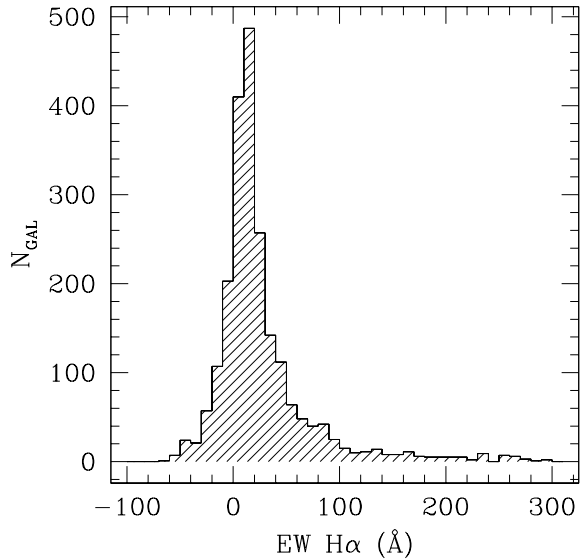}
\begin{footnotesize}
\caption{The distribution of equivalent widths of H$\alpha$
for the whole sample showing the asymmetry towards larger EW.
}
\end{footnotesize}
\label{EWhisto}
\end{figure}


The measurement of the background correction and the choice of the 2.5$\mathfrak{c}$ 
selection are illustrated in
Figures~\ref{Fcuts}c and d. Higher redshift galaxies could have emission lines in the 
817 nm filter, \eg $\lambda$5007 at $z = 0.63$.
Three fields of total area 111.6 arcmin$^2$ displaced about 13 arcmin
from the cluster centre were summed.
A 2$\mathfrak{c}$ cut eliminates fewer objects, 3$\mathfrak{c}$ removes more sources, 
and the
$2.5\mathfrak{c}$ curves remove background sources well as seen in Figure~\ref{Fcuts}c. 
Consequently, the number of detected objects is 
taken to lie between these limits as an estimate of the choice for pruning.

Figures~\ref{Fcuts}b and 2d show the objects remaining with the condition 
that $F_{817} \ge 2.5\mathfrak{c}$.
In all, 2192 objects were detected near the centre of Abell 2465 and
222 emission line objects remained after pruning. In Figure~\ref{Fcuts}a, 137 objects lie
below the lower $2.5\mathfrak{c}$ curve. This is more than expected from a Gaussian error
distribution and these are expected to represent absorption line sources in the sample.

The spectroscopy in Paper I is neither as extensive nor reaches as faint
as the present photometric determinations, 22 measurements of the equivalent 
width of H$\alpha$ can be compared in Figure~\ref{EWcompare}. The photometric
measures tend to be the larger and a least squares fit gives
$EW_{spec} = 0.76(EW_{phot}-0.08)$. This empirical
correction is considered due to the simplifications in deriving the equivalent widths.
This and galactic extinction are 
not applied to the measures in Tables~\ref{1.tab}
and \ref{2.tab}, but are used in the SFR estimates in Table~\ref{3.tab}.
\begin{figure}
\includegraphics[width=8cm]{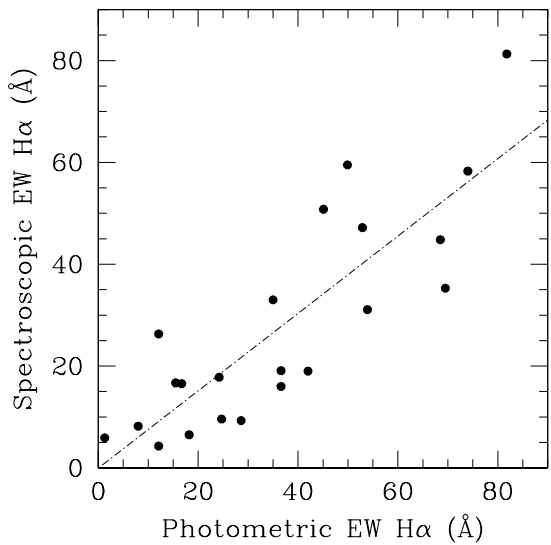}
\begin{footnotesize}
\caption{Comparison of the 22 H$\alpha$ equivalent widths measured
spectroscopically in Paper I and photometrically in this paper.
The dashed line has slope 0.76.
}
\end{footnotesize}
\label{EWcompare}
\end{figure}

Table~\ref{1.tab} presents the measurements of the galaxies described. 
Columns 1 and 2 give the measured
(J2000) coordinates. Columns 3 - 6  are the measured $i_g$ and $m_{817}$ filter magnitudes
and their errors from the {\sc Sextractor} program.
Column 7 and 8  are the resulting equivalent widths of H$\alpha$ and
their errors in~\AA~ 
according to equation 2, positive values 
denote emission and negative absorption. Column 6 is H$\alpha$ flux
multiplied by $10^{16}$ in ergs$^{-1}$cm$^{-2}$arcmin$^{-2}$ units.

\begin{table*}
\caption{The observed photometric quantities measured in Abell 2465}
\label{1.tab}
\begin{small}
\begin{tabular}{rrrrrrrrrr}
\hline

$\alpha_{J2000}$ &
$\delta_{J2000}$ &
$i_g$&$\varepsilon_{i_g}$&
$m_{817}$&$\varepsilon_{817}$&
EW H$\alpha$ &$\varepsilon_{\rm{H}\alpha}$&
$F_{\rm{H}\alpha}$&$\varepsilon_{F_{\rm{H}\alpha}}$
\\
(1)&
(2)&
(3)&
(4)&
(5)&
(6)&(7)&(8)&(9)&(10)\\
\hline
    339.68359 &     -5.76283 &  20.004 &$\pm$   0.011 &  19.238 &$\pm$   0.046 &  100.5 &$\pm$   15.8 &  44.38 &$\pm$   4.82 \\
    339.68852 &     -5.77682 &  22.126 &$\pm$   0.023 &  22.704 &$\pm$   0.298 &  -19.1 &$\pm$    9.4 &  -1.31 &$\pm$   1.31 \\
    339.69149 &     -5.85340 &  22.039 &$\pm$   0.042 &  24.003 &$\pm$   1.799 &  -50.6 &$\pm$  102.8 &  -3.84 &$\pm$   2.43 \\
    339.69190 &     -5.91493 &  20.835 &$\pm$   0.022 &  21.010 &$\pm$   0.183 &    1.3 &$\pm$    0.5 &   0.29 &$\pm$   3.89 \\
    339.69195 &     -5.92532 &  18.769 &$\pm$   0.008 &  18.962 &$\pm$   0.071 &    0.2 &$\pm$    0.0 &   0.25 &$\pm$   9.90 \\
    339.69220 &     -5.89481 &  22.493 &$\pm$   0.065 &  21.521 &$\pm$   0.187 &  140.1 &$\pm$  118.5 &   6.08 &$\pm$   2.48 \\
    339.69225 &     -5.87150 &  22.284 &$\pm$   0.038 &  22.619 &$\pm$   0.366 &   -7.8 &$\pm$    5.4 &  -0.46 &$\pm$   1.77 \\
    339.69318 &     -5.80899 &  21.657 &$\pm$   0.017 &  21.209 &$\pm$   0.087 &   55.2 &$\pm$   13.9 &   5.50 &$\pm$   1.52 \\
    339.69322 &     -5.89046 &  19.927 &$\pm$   0.008 &  20.260 &$\pm$   0.072 &   -7.7 &$\pm$    1.1 &  -3.96 &$\pm$   3.07 \\
    339.69352 &     -5.67319 &  19.771 &$\pm$   0.008 &  19.749 &$\pm$   0.054 &   14.7 &$\pm$    1.8 &   8.57 &$\pm$   3.65 \\

\hline
\end{tabular}
\end{small}

\begin{tabular}{l}
{This table is available in it entirety in machine readable and Virtual Observatory (VO)
forms in the online journal. 
A portion is shown here for guidance regarding its form and content.
}\\
\end{tabular}
\end{table*}


\section{The H$\alpha$ distribution and star forming rate}
The SFR rate is obtained from Kennicutt's (1998) H$\alpha$ calibration.
which is sensitive to absorption. Comparisons between H$\alpha$
and [OII] and infrared measurements which are less sensitive to reddening
indicate that H$\alpha$ does not produce systematic errors, but has larger
errors, typically varying by $\pm$~15\% to
30\% (Moustakas \etal 2006).
The distributions of the H$\alpha$ galaxies can be compared with those found for
continuum sources.

\subsection{The H$\alpha$ star forming rate in Abell 2465}
The locations of H$\alpha$ emission sources on the sky 
using the 2.5$\mathfrak{c}$ cut are in 
Figures~\ref{Hemsplot.eps} and \ref{contours.ps}. 
Due to the double structure of Abell 2465, the subclusters 
are considered separately and combined for the whole cluster.
Both SW and NE components of Abell 2465, have $R_{200} = 1.2$~Mpc (Paper I), the 
5.2 arcmin circles in Figure~\ref{Hemsplot.eps}
centred on the brightest cluster galaxies (BCGs) also lie near the X-ray centres.
For the whole cluster, $M_{cl}$ and $R_{200}$ were defined using standard formulae
(\eg Carlberg, Yee, \& Ellingson 1997; Koyama \etal 2010),
$R_{200} = 2.47(\sigma/1000)/\sqrt{\Omega_\lambda + \Omega_0(1+z)^3}$ Mpc and
$M_{cl} = 1.71 \times 10^{15}(\sigma/1000)^3/\sqrt{\Omega_\lambda + \Omega_0(1+z)^3} M_\odot.$
The redshift data in Paper I give  $\sigma = 919$ \kms, so $R_{200} = 2.0$ Mpc and 
$M_{cl} = 12 \times 10^{14} M_\odot,$ shown as the outer dotted circle centred midway 
between the two BCGs in Figure~\ref{Hemsplot.eps}. In the following, $M_{cl}$ and $M_{200}$ will be used interchangebly.

The  Figure~\ref{contours.ps} isophotes were
constructed from the $F_{H \alpha}$ 
using adaptive kernal smoothing 
(Silverman 1986) and compare with the $i'$ continuum luminosity in Paper I.
The $H\alpha$ emission seems displaced from the continuum and X-ray peaks, but closer
to the NE component than the SW component. 

The three samples derived from the $\mathfrak{c}$ cuts, as 
emission line galaxies located in Figure~\ref{Fcuts} using 2, 2.5, 
and 3 $\mathfrak{c}$ cuts was used to
estimate the range in the total H$\alpha$ flux,
$\mathcal{L}_{H\alpha} = \sum F_{\rm{H}\alpha}$ from the selection method. Total 
values inside $R_{200}$ were derived for the SW and NE
and for the whole cluster defined above,
after subtracting the background scaled to the same solid angles.

\begin{figure}
\subfloat{\includegraphics[width=0.5\textwidth]{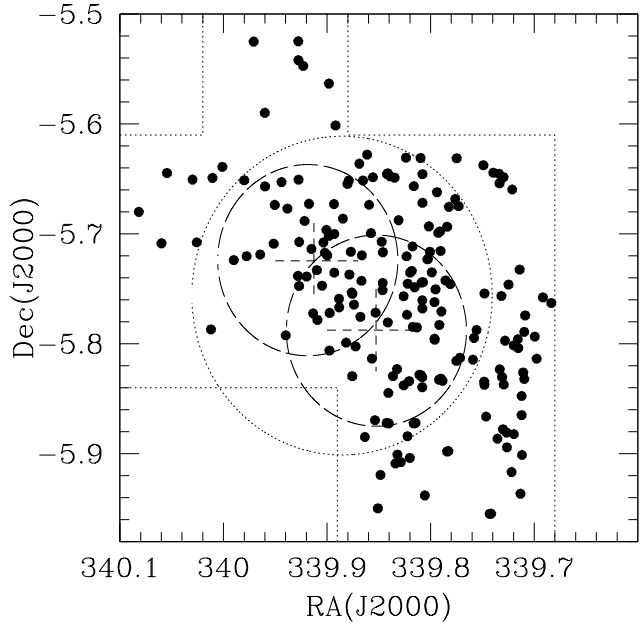}}
\begin{footnotesize}
\caption{Positions of galaxies with detected H$\alpha$ emission in
the central region of Abell 2465 shown in Figure~\ref{Fcuts}. The dashed crosses at the 
centres of the circles with radii $R_{200} \approx 1.2$~Mpc lie on the BCGs of
each clump. The outer dotted circle is centred half way between the two subclusters with
radius $R_{200} = 2.0$ Mpc representative of the whole cluster. Horizontal and vertical dotted 
lines represent the limits of the area scanned for H$\alpha$.
}
\end{footnotesize}
\label{Hemsplot.eps}
\end{figure}
\begin{figure}
\subfloat{\includegraphics[width=0.45\textwidth]{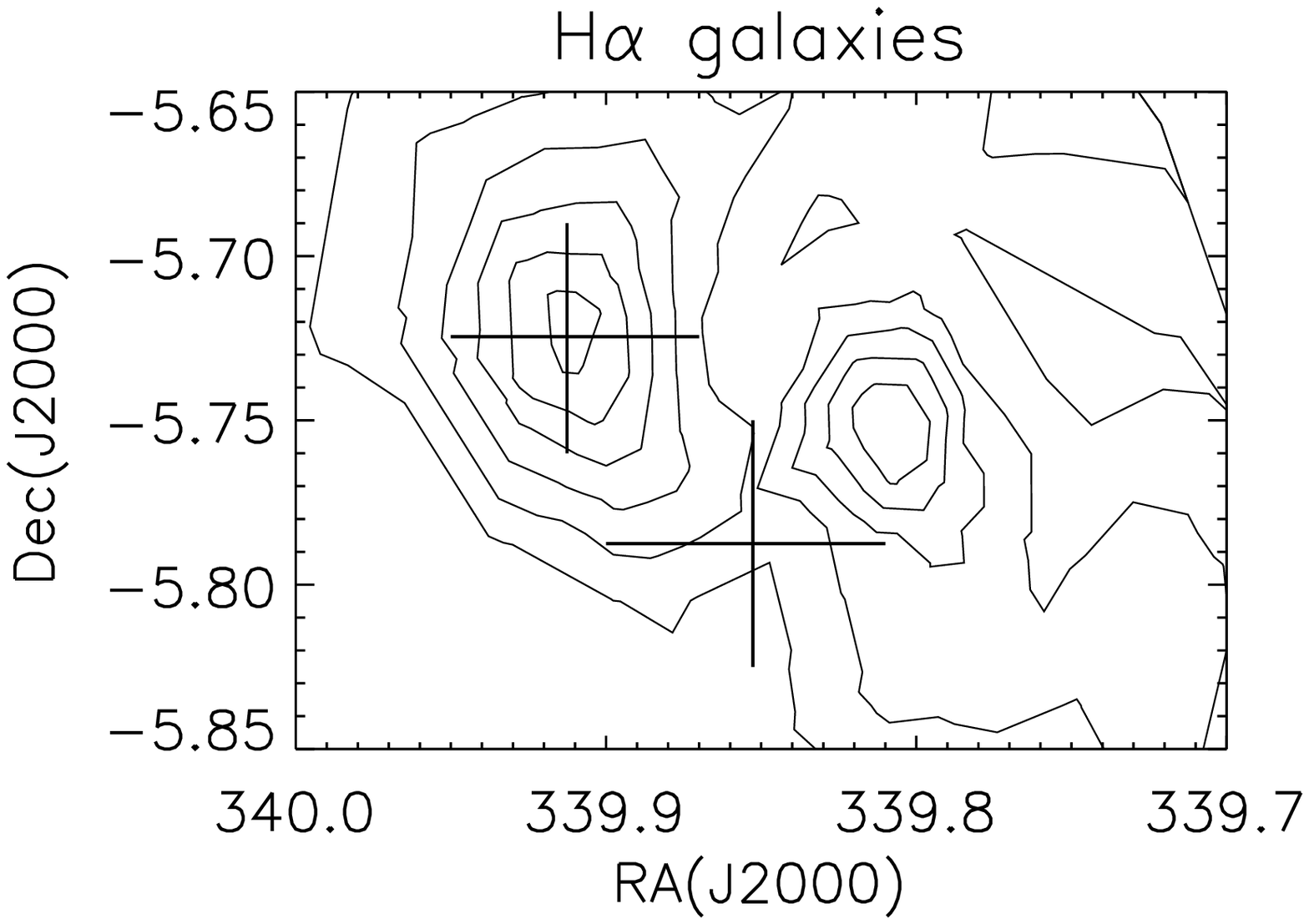}}

\subfloat{\includegraphics[width=0.45\textwidth]{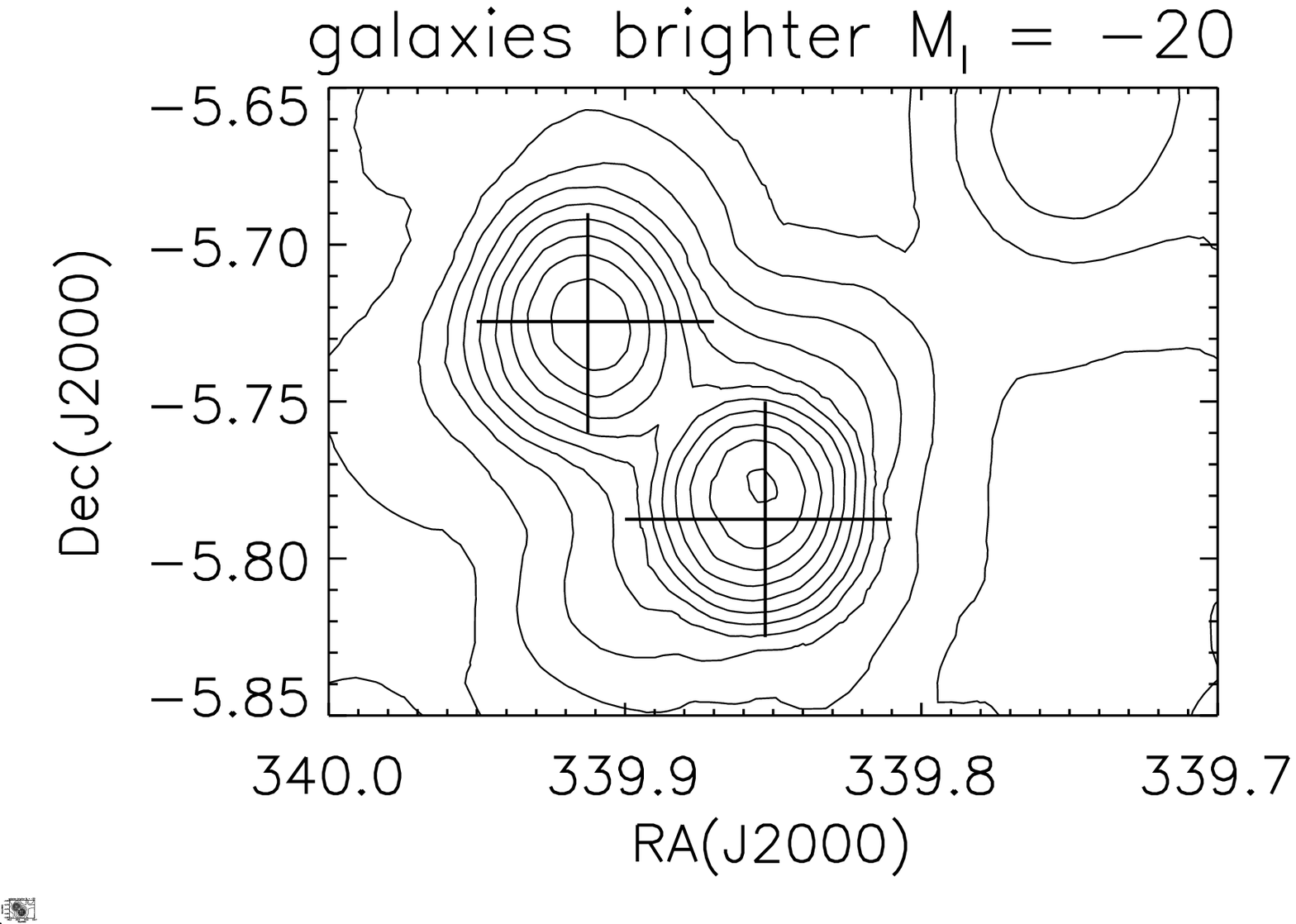}}

\subfloat{\includegraphics[width=0.45\textwidth]{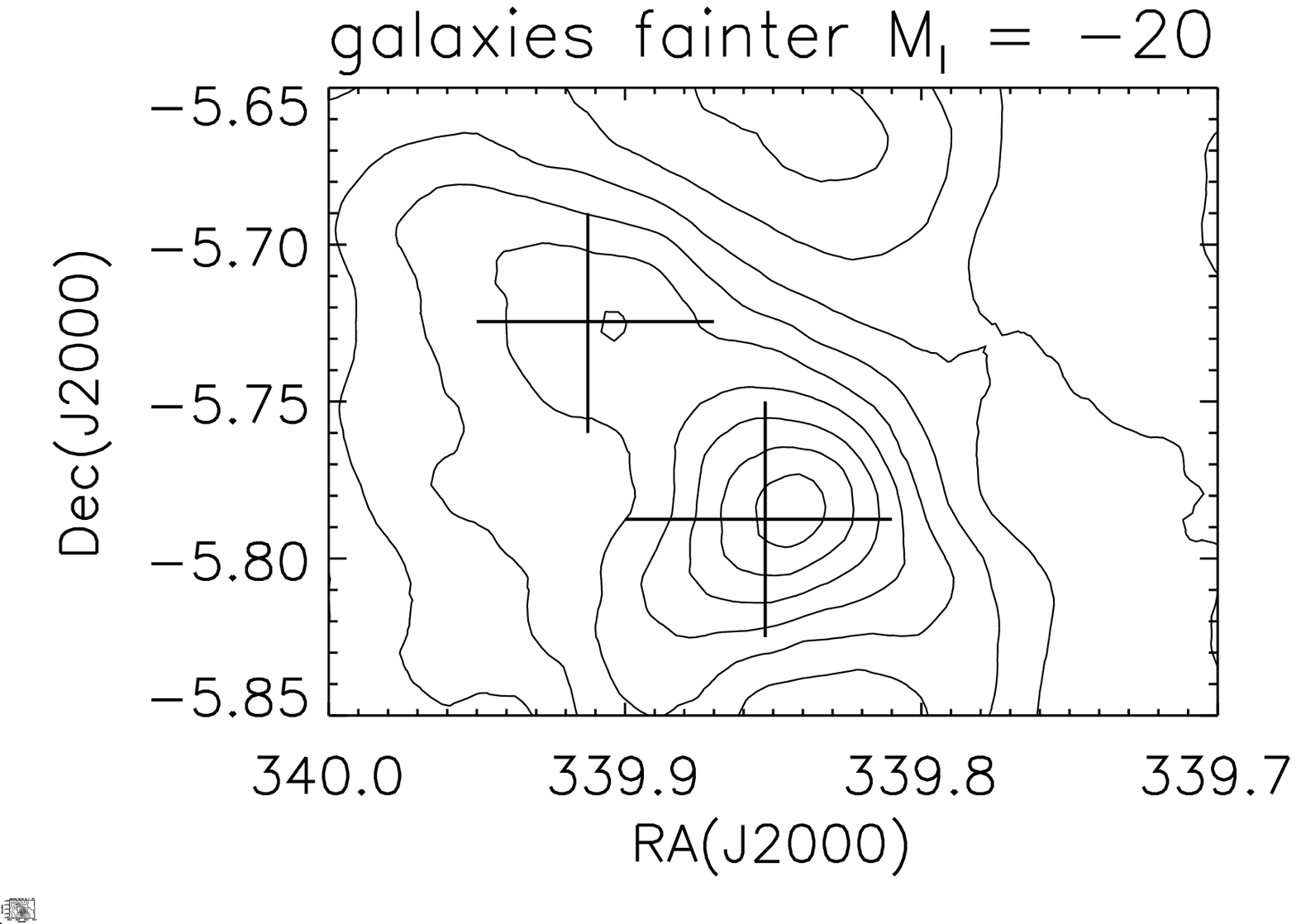}}
\begin{footnotesize}
\caption{(Top) H$\alpha$ emission isophotes in
the central region of Abell 2465 shown in Figure~\ref{Hemsplot.eps}. 
(Bottom) The isophotes in the two panels are from the 
continuum $i'$ data in Paper I and divides the galaxies at $M_I$ = -20.0 mag.
The crosses lie on the BCGs of each clump as before. 
}
\end{footnotesize}
\label{contours.ps}
\end{figure}

Using Table~\ref{2.tab}, the SFR for each entry was calculated,
taking the luminosity distance to Abell 2465 as 1224 Mpc and the 817 nm galactic 
extinction as 0.08 mag. from NED.
Including the cosmological correction, $(1+z)^{-1}$ to the equivalent widths, 
the absolute luminosity of H$\alpha$ is: 
$$ L_{H\alpha} = 1.55 \times 10^{56} \mathcal{L}_{H\alpha} 
\rm{~ergs~s^{-1}~cm^{-2}}.$$

The luminosity function (LF) of $L_{H\alpha}$, constructed from 
binning the data is in Figure~\ref{HaLF}; errors bars are Poisson errors
from the number counts and an internal reddening correction for the galaxies is not applied. 
Bai \etal (2009) and others, \eg Hwang \etal (2012a) show that the Schecheter (1976) 
function, 
$$\Phi(L) = \Big(\frac{\Phi^*}{L^*}\Big)\Big( \frac{L}{L^*}\Big)^\alpha \exp(-L/L^*).$$
fits the starforming LF.
Bai \etal (2009) showed that a fixed $\alpha = -1.41$ fits a number of
galaxy clusters well. Adopting $\alpha = -1.4$ gives $\log L^* = 42.5$ 
in Figure~\ref{WISE_SCHECT.eps}. Used as a
completeness test of the H$\alpha$ data, it shows a
low luminosity cutoff limit, $\log L/L^* = -1.2,$ which corresponds to
SFR = 1.5 M$_\odot$ yr$^{-1},$ or $L/L^* = 0.063$

For the $\Phi(L)$ above, the total luminosity density is 
the well-known relation with the $\Gamma$-finction, $L_{tot} = \Phi^*L^*\Gamma(2+\alpha)$,
where $\Gamma(0.6) = 1.49.$

However, for a lower non-zero cutoff in $L$, $x$, the total luminosity is:
$$L_{x} = \int ^{\infty}_{x}dLL\Phi(L) = \Phi^*L^*\Gamma(\alpha + 2, x),$$
where $\Gamma(2 + \alpha, x)$ is the incomplete gamma function. 

Knowing $x, L_{x},$ and $\alpha$ gives $L^*\Phi^*$ and the observed $L_{x}$ can be 
corrected for incompleteness. Usually
a low luminosity cutoff near 0.3 $M_\odot yr^{-1}$ is chosen (\eg Bai \etal 2009).
For the H$\alpha$ data this corresponds to $L/L^* = 0.012$. so the
corrected luminosity is $$L_{0.3} = L_x\Gamma(0.6,0.012)/\Gamma(0.6,0.063),$$
hence $L_{0.3} = 1.2L_{x}.$

Table~\ref{2.tab} shows the H$\alpha$ flux measurements. 
In the summation of the fluxes, all sources above the cutoff in Figure~\ref{HaLF}
were included.
Column 1 is the $\mathfrak{c}$ cut sample,
Columns 2-5 give $\mathcal{L}_{H\alpha}$ and their estimated jackknife errors, for each 
region in $10^{-13}$erg~s$^{-1}$cm$^{-2}$ units
and Column 5 is the background determination in units of
$10^{-15}$erg~s$^{-1}$~cm$^{-2}$arcmin$^{-2}$.

\begin{figure}
\includegraphics[width=8cm]{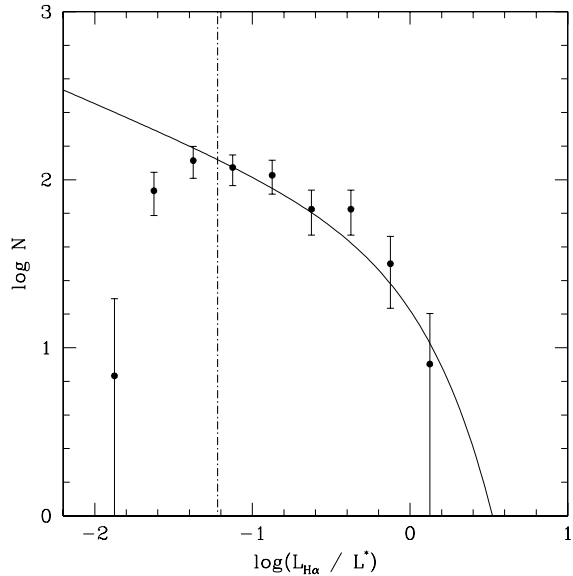}
\begin{footnotesize}
\caption{The H$\alpha$ luminosity function of Abell 2465 within 
$R_{200} = 2.0 $ Mpc. Data are fitted to $\Phi(L)$ for $\alpha = -1.4$ 
and $\log L^* = 42.5$.
Error bars are Poisson errors from the number counts. The estimated low luminosity
cutoff is shown as the vertical dash-dot line.
}
\end{footnotesize}
\label{HaLF}
\end{figure}


\begin{table}
\caption{H$\alpha$ fluxes $\mathcal{L}_{H\alpha}$ for Abell 2465
(erg~s$^{-1}$~cm$^{-2}$)
}
\label{2.tab}
\begin{center}
\leavevmode
\begin{tabular}{lcccc}
\hline
      Sample&WHOLE&SW&NE&Background\\
 &$\cdot$10$^{13}$ & $\cdot$10$^{13}$& $\cdot$10$^{13}$&$\cdot$10$^{15}$ arcmin$^{-2}$\\
     (1)&(2)&(3)&(4)&(5)\\
\hline
2$\mathfrak{c}$&3.1$\pm$0.6&1.2$\pm$0.2&1.5$\pm$0.6&0.3$\pm$0.1\\
2.5$\mathfrak{c}$&2.5$\pm$0.5&1.0$\pm$0.2&1.25$\pm$0.6&0.25$\pm$0.1\\
3$\mathfrak{c}$&1.85$\pm$0.4&0.8$\pm$0.2&0.8$\pm$0.3&0.2$\pm$0.1\\

\hline
\end{tabular}
\end{center}
\end{table}


The SFR calculated using Kennicutt (1998),
$SFR =  L_{H\alpha}/1.26 \times 10^{41} M_\odot$~yr$^{-1}$, also 
requires a correction for the internal reddening of the galaxies.
For the 42 galaxies in Paper I with emission, 
$<L_{H \alpha}/L_{H_\beta}>~ = 4.23$ and the standard interstellar extinction 
curve (Osterbrock 1989)
gives the mean extinction for H$\alpha$ of $A_{H_\alpha} = 0.9$~mag., similar to
the 1.1 mag. in Kennicutt (1998).
Consequently a 1.0 mag. correction was made along with the 0.76
factor in Section 2.3 and the 1.2 factor found above, giving a 2.29 multiplicative
correction which was used to get the final values of SFR in Table~\ref{3.tab},
where Column 1 gives the cut sample, Columns 2-4 give SFR 
and their errors in $M_\odot$ yr$^{-1}$ for the different regions.

\begin{table}
\caption{Star formation rates in Abell 2465 from H$\alpha$ ($M_\odot$~yr$^{-1}$)
}
\label{3.tab}
\begin{center}
\leavevmode
\begin{tabular}{lcccc}
\hline
      Sample&SW&NE&WHOLE\\
     (1)&(2)&(3)&(4))\\
\hline
2$\mathfrak{c}$&276$\pm$48&364$\pm$161&$665\pm$180\\
2.5$\mathfrak{c}$&217$\pm$62&290$\pm$158&533$\pm$163\\
3$\mathfrak{c}$&175$\pm$62&197$\pm$99&412$\pm$116\\
$M/10^{14}M_\odot$&4&4&12& \\
\hline
\end{tabular}
\end{center}
\end{table}


The spectroscopic H$\alpha$ measurements 
of galaxies in Table 3 of Paper I give a lower limit to the SFR. They are 
an incomplete sample of cluster galaxies brighter than $i' \approx +20$.
The sum of the H$\alpha$ fluxes for the SW + NE components inside
their $R_{200} = 1.2$~Mpc radii is $4 \times 10^{-14}$  ergs~sec$^{-1}$~cm$^{-2}$, or
$\Sigma SFR~\ga~123~M_{\odot}$yr$^{-1}$ for 17 spectroscopically verified cluster
members inside the 1.2 Mpc circles in 
Figure~\ref{Hemsplot.eps}.

\section{Star formation rate from WISE satellite data}
WISE satellite data (Wright \etal 2010; Jarrett \etal 2011; 
and {\it WISE Explanatory Supplement}
{\footnote{http://wise2.ipac.caltech.edu/docs/release/allsky/expsup/}})
provide an independent determination of the SFR from infrared 
dust emission. Stern \etal (2012), Chung \etal (2011), Hwang \etal (2012a), 
and others have discussed
these data for galaxy clusters. At $z =0.245$, many cluster members in
Abell 2465 are near the limit of WISE and few objects are detected with its 
22~$\mu$m~$[w4]$ band, so the 12~$\mu$m~$[w3]$ and shorter wavelength 
$[w2]$ and $[w1]$ bands were employed. 
Objects with $[w3]$ magnitudes 5$\sigma$ or better were selected. Following 
Jarrett \etal (2011) and Stern \etal (2012), star 
forming galaxies were chosen with
$[w1] - [w2] \le 0.7$ and $2.7 \le [w2] - [w3] \le 5.0$ to eliminate AGNs. 
The total flux using $[w3]$ was calculated as in Hwang \etal (2012a)
employing the zero point
calibration in Jarrett \etal (2011), 31.674 Jy for a 0 mag. Vega source.
The total infrared dust luminosity, $L_{IR}$, was
estimated from the Chary \& Elbaz (2001) templates 
employing a top hat response, using 11.56~$\mu$ m 
for central wavelength and 5.51~$\mu$m resolution 
(Jarrett \etal 2011). 
A ring of radii 18-28 arcmin centred on the cluster was used to estimate the background.
The SFR was
calculated using $$SFR = 1.72 \times 10^{-10}(L_{IR}/L_{\odot})~ M_{\odot}~\rm{yr}^{-1}$$ 
which is Kennicutt's (1998) relation for a Salpeter 
initial mass function (IMF). Chen \etal (2013) note that 
the Chabrier (2003) initial mass function lowers SFR by about 1.6. Hwang \etal (2012a, b)
show that the SFRs from the spectral energy distribution 
fitting to the WISE flux densities of galaxies agree with those from optical spectra,

The IR luminosity function for the whole cluster
is shown in Figure~\ref{WISE_SCHECT.eps} fitted with $\alpha = -1.4$ as in Section 3, with 
$\log L^* = 10.5.$ The cutoff corresponding to the $5\sigma$ selection limit is at
SFR = 2.5 M$_\odot$yr$^{-1}$ which is shown in the diagram and corresponds to
$x = L/L^* = 0.46$  

\begin{figure}
\includegraphics[width=8cm]{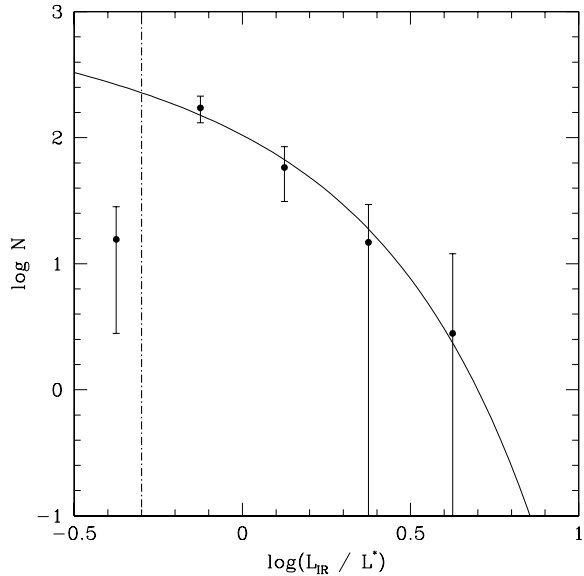}
\begin{footnotesize}
\caption{The IR luminosity function of the whole cluster sample of Abell 2465 within 
$R_{200} = 2.0 $ Mpc. Data are fitted to $\Phi(L)$ with $\alpha = -1.4$ 
and $\log L^* = 10.5$. Error bars are Poisson errors from the number counts.
The estimated completeness limit is the vertical dash-dot line.
}
\end{footnotesize}
\label{WISE_SCHECT.eps}
\end{figure}
The binned
number counts and Poisson errors is in Figure~\ref{WISE_SCHECT.eps} where 
there is a cutoff
below $\log L_{IR}/L^* \approx -0.3$, which corresponds to 
$SFR \approx 3 M_\odot$~yr$^{-1}$ at Abell 2465. 
The total $SFR$ to 0.3 $M_\odot$~yr$^{-1}$, for comparison
with that of H$\alpha$ and other clusters in the literature, 
is again calculated using:
$$ L_{0.3} = \Gamma(0.6, 0.055)/\Gamma(0.6, 0.51)L_{IR} = 2.15$$
The SFR using this method for the same three cluster divisions as defined for H$\alpha$
are in Table~\ref{WISESFR}.  
Considering the observational
uncertainties of the two methods, the agreement between 
Tables~\ref{3.tab} and ~\ref{WISESFR}, which is better than a factor of two, seems satisfactory.

\begin{table}
\caption{Star formation rates in Abell 2465 from WISE data }
\label{WISESFR}
\begin{center}
\leavevmode
\begin{tabular}{lcccc}
\hline
      Sample&SW&NE&WHOLE\\
     (1)&(2)&(3)&(4))\\
\hline
SFR (M$_\odot$yr$^{-1})$&206$\pm$75&445$\pm$142&890$\pm$114\\
No objects&42&52&127 \\
\hline
\end{tabular}
\end{center}
\end{table}


\section{Comparison with other galaxy clusters}
\subsection{Projected distribution of the SF galaxies}
The galaxy counts herein have been made inside \R200 and for SFR $\ge 0.3$ \MSYR, but
many authors employ 0.5\R200 and alternate cutoffs for SFR. As shown
above, a correction for the different
SFR cutoffs can be made assuming the form of $\Phi(L)$, but a conversion for the different
choise of radius depends on how the SFR changes within the cluster, and as suggested by many
investigators, it may not obey the mass profile. Figure~\ref{radial_SFR} shows the
accumulated projected sum of the H$\alpha$ and IR SFR in Abell 2465, 
after background subtraction, 
within a given radius for the cluster as a whole and for the SW and NE sub-clusters. Although 
the H$\alpha$ SFR in the SW region appears weakly more concentrated, the star
forming galaxies in Abell 2465 do not follow the NFW profile found for the $i$-band light
in Paper I. Defining the factor $\mathfrak{K}$ to be the ratio of projected galaxies
inside 0.5\R200 to those inside \R200,
$$\mathfrak{K} = \sum SFR (0.5R_{200})/\sum SFR (R_{200}) \approx 0.4.$$
This result seems to confirm Chung \etal (2011) using WISE and Webb \etal (2013) 
employing Spitzer Multiband Imaging Photometer for Spitzer (MIPS) data. 
For an NFW profile with a concentration parameter $c \approx 6,$ $\mathfrak{K} = 0.6,$
({\L}okas \& Mamon 2001) and for a uniformly filled circle, $\mathfrak{K} = 0.25,$ This
agrees with the finding that star forming galaxies are less concentrated to the centres 
of their clusters (\eg Rines \etal 2005, Koyama \etal 2010; Webb \etal 2013; Muzzin \etal 2012).

\subsection{Position of Abell 2465 in the redshift and mass relations}
Studies of the SFR in galaxy clusters employing optical and infrared data
over a range of redshifts include Finn \etal (2005), Bai \etal (2007; 2009),
Koyama \etal (2010), Chung \etal (2011), Biviano \etal (2011), Shim \etal (2011),
Popesso \etal (2012), and Webb \etal (2013). The total SFR normalized 
by the cluster mass, $\Sigma SFR/M_{cl}$ has been compared with redshift, $z$ 
and total $M_{cl}$. There is considerable scatter in these relations; often clusters without
marked substructure are mixed with interacting ones.

>From Tables~\ref{3.tab} and ~\ref{WISESFR}, $\Sigma SFR/M_{cl}$ in Abell 2465 is
plotted against $z$ and $M_{cl}$ in Figure~\ref{SFRzM}. Several studies, following Finn \etal
(2005), have found relations between $\Sigma SFR/M_{cl}$ and redshift: $(1 + z)^6$
(Koyama \etal 2010), $(1 + z)^{5.3}$ (Bai \etal 2007; 2009) with similar results by
Biviano \etal (2011) and Popesso \etal (2012) who obtained a $z^{1.77}$ relation.
These empirical functions 
generally give similar fits to the data when allowance is made for their different
radial and SFR cutoffs. Figure~\ref{SFRzM}a plots the components of Abell 2465 with the 
$(1 + z)^{5.4}$ relation of Webb \etal (2013) which uses galaxies within \R200, but has been
corrected to account for their 3 \MSYR cutoff compared to 0.3 \MSYR in this
paper. In this case, both sub-components of Abell 2465 and the cluster taken as a whole lie
above the mean relation by a factor of $\sim7$.

For the mass dependence, the situation is more complicated. The scatter is larger and $M_{cl}$
is less well determined compared to $z$. Also $\Sigma SFR/M_{cl}$ must be converted for redshift
as well as cutoff. Finn \etal (2005) suggested an inverse relation between SFR and $M_{cl}$
and Bai \etal (2007) found a $M_{cl}^{-0.9}$ decline. Figure~\ref{SFRzM}b shows the Abell 2465
data with the $M_{cl}^{-1.5}$ curve from Webb \etal (2013) which has been converted from
$z = 0.4$ and SFR $ > 3$~\MSYR~to $z = 0.245$ and 0.3 \MSYR. The SFR for all components is
at least an order of magnitude above the mean relation.

Although the scatters in the $\Sigma SFR/M_{cl}$ relations are large, Abell 2465 indicates
a higher SFR for its redshift and mass.  For comparison, three additional galaxy 
clusters are plotted in 
Figure~\ref{SFRzM}. For details on other galaxy clusters, the reader is referred to the
references above. The Bullet Cluster (Chung \etal 2010) and Cl0024+1654 (Bai \etal 2007;
Umetsu \etal 2010) are both considered merging systems and also show enhanced SFR. By contrast,
the Coma Cluster (\eg Finn \etal 2005; Bai \etal 2007) is a relatively symmetrical and quiet
galaxy cluster and it lies lower in Figure~\ref{SFRzM}.
This adds to the evidence that interacting clusters can have a higher SFR and
suggests that a cluster's dynamical state 
cluster is an additional parameter in these plots.

\begin{figure}
\includegraphics[width=6cm]{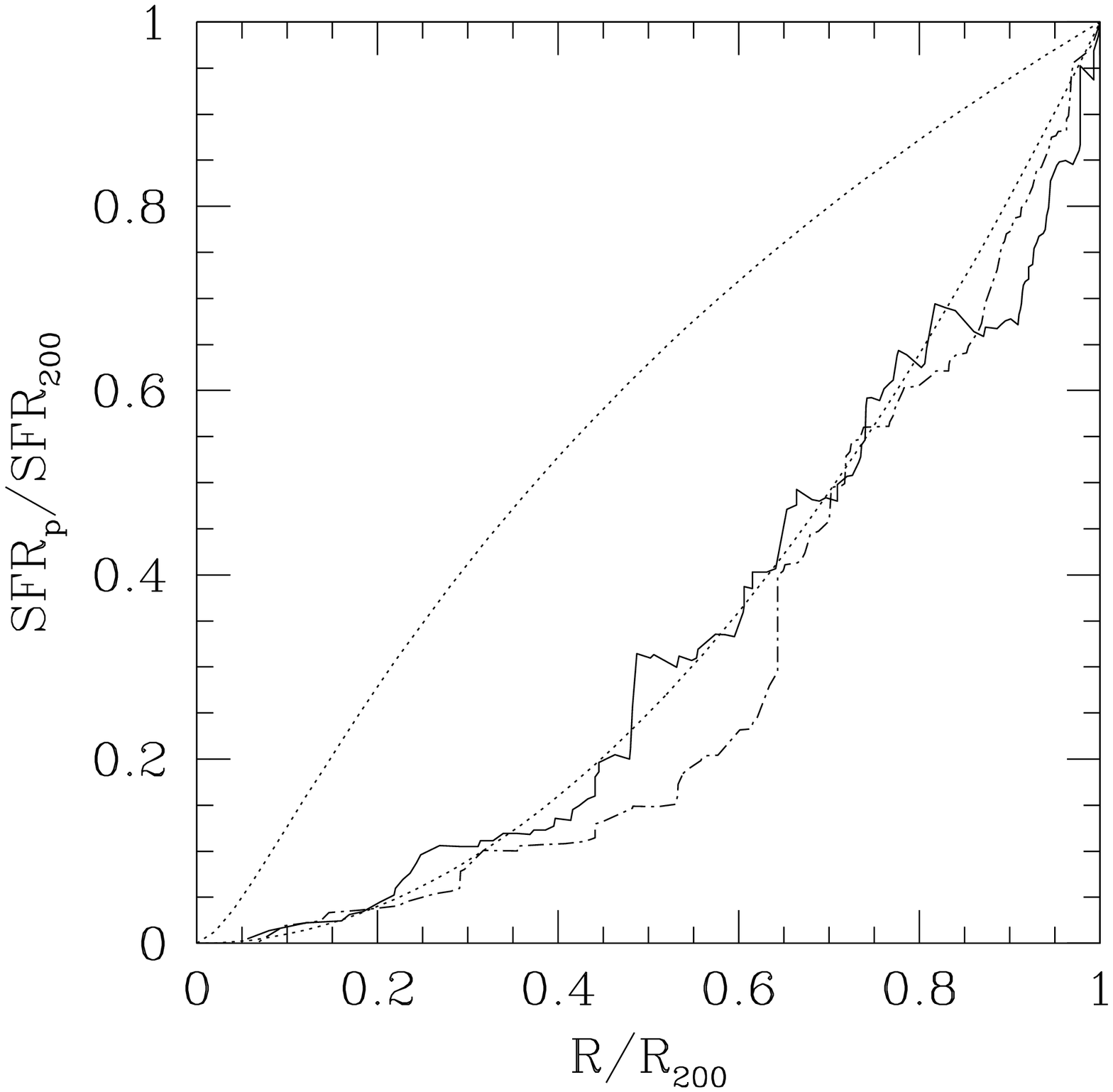}
\includegraphics[width=6cm]{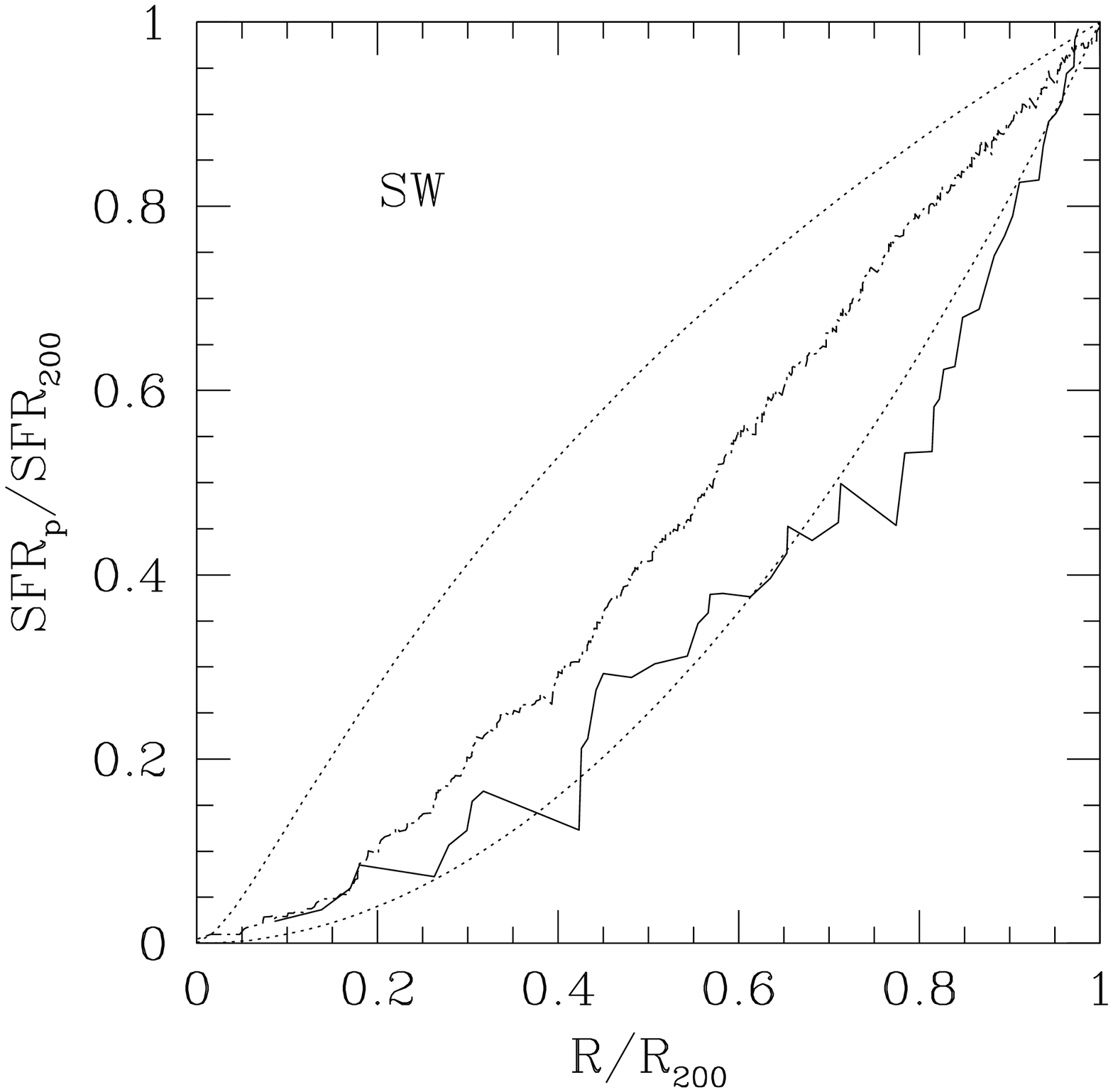}
\includegraphics[width=6cm]{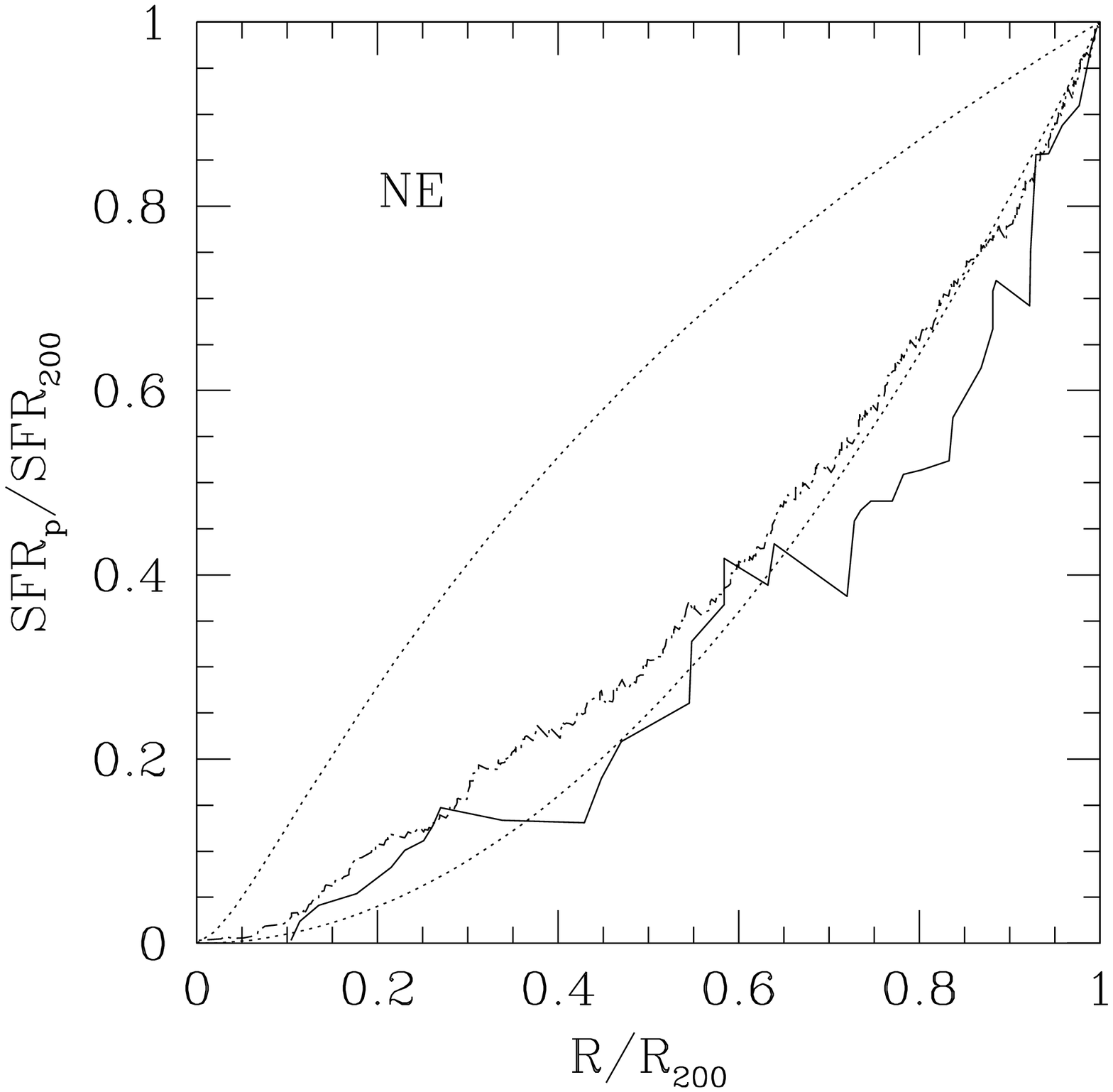}
\begin{footnotesize}
\caption{The projected SFR rate surface densities in Abell 2465 compared with the 
projected mass surface densities for a $c = 6$ NFW profile and a uniform surface distribution
(upper and lower dotted curves). The IR data follow the solid curve and the H$\alpha$ are shown
as the dot-dashed relation. The top curve is for the whole cluster and the lower panels present 
data for the SW and NE regions. 
}
\end{footnotesize}
\label{radial_SFR}
\end{figure}
\section{Morphologies of the strongest H$\alpha$ sources}
Thirty strong H$\alpha$ emission line sources 
in Table~\ref{1.tab} are shown in 
Figure~\ref{empictures} in 20 arcsec square (77 kpc) sections. These are from the
$i$ CFHT image described in Paper I with FWHM = 0.47 arcsec seeing. 
Figure~\ref{WISEpictures} is similarly a sample of 25 among the brightest IR WISE sources 
found here. Although a
census of morphological types for the smaller galaxies in Abell 2465 
using the currently available resolution seems outside the scope of this paper,
many of these H$\alpha$ sources appear to be interacting or have a disturbed appearance. 
Examples in Figure~\ref{empictures} include 8, 11, 12, 14, 18, 19, 25, and 30 which may be
multiple mergers. Further binary mergers could include the
systems 6, 9, 10, 13, 17, 23, 27, and 29,
while 1, 5, 7, 15, 16, 20, and 24 appear disturbed and asymmetrical. 

For WISE IR sources
without redshifts, the interpretation is more difficult. Although
for the H$\alpha$ sources, the likelihood that a bright redshifted line, \eg $\lambda$5077
at $z = 0.63$ would fall in the bandpass of the H$\alpha$ filter is under a few percent
(\eg Kellar \etal 2012), the distances to the IR sources is unrestrained and 
although a background
correction was made for the whole sample, individual galaxies could be in the fore- or 
background. As a result, all galaxies in, except number 7, in Figure~\ref{WISEpictures} were 
chosen to be either H$\alpha$ detections or spectroscopic redshifts placing them in Abell 2465.
Objects that appear interacting (4, 7, 9, 11, 14), have a
possible companion (1, 2, 3, 5, 21, 23) or seem disturbed (2, 3, 13, 18, 20, 24)
in some way. 

Many of the  H$\alpha$ and IR sources are not detected congruently. This appears to be 
because many of them are faint and close to the detection limits of the H$\alpha$ survey
and WISE which are different.
The H$\alpha$ data reach fainter sources than
WISE IR ($\sim 1.5$~M$_\odot$yr$^{-1}$ compared to 
$\sim 2.5~$M$_\odot$yr$^{-1}$). On the one hand, objects detected only in H$\alpha$
would have weaker IR, but on the other hand, sources found only in the IR are likely
to be very dusty and hide the H$\alpha$ and shorter wavelengths of light.
This is indicated from comparisons between the H$\alpha$ and IR SFR which show that the IR
can give the higher value (\eg Hopkins \etal 2001; Bai \etal 2007), although in general
H$\alpha$ and IR flux densities correlate well (Kennicutt \etal 2009).

\section{The $U$ and $B$ observations}
Abell 2465 was observed 2009 October 14-18 through Johnson $U~\rm{and}~B$ 
filters with the 4K
CCD camera\footnote{Details of this instrument can be found at: 
http://www.astronomy.ohio-state.edu/MDM/MDM4K}
attached to the 1.3 m telescope at the MDM Observatory.
This device has 4064$^2$ pixels and covers 21.3 arcmin 
$\times$ 21.3 arcmin. The gain was 2.3 ADU with a readout noise of 5 e$^-$.
At readout,
2 $\times$ 2 pixel binning was employed yielding 0.63
arcsec per pixel. All images were inspected to reject guiding jumps, a
problem with the 1.3 m telescope, and 45 ten minute exposures of
the cluster in $U$ and 30 in $B$ were secured 
at airmasses less than 1.5, yielding total
summed integrations of 27000 s and 18000 s in $U$ and $B$ respectively. Stars
in both $U$ and $B$ final images have FWHM = 2.1 arcsec. Data reductions used
IRAF as above, viz. bias subtraction, flatfielding from dome flats,
and cosmic ray rejection. 

The images were calibrated to the Johnson
system using standard stars in the SA92 field (Landolt 1992) 
observed with airmasses of 1.17 to 2.06 on the
photometric nights of 2009 October 15, 16, and 17 giving extinction
coefficients of $a_U = 0.445$ and $a_B = 0.217.$ A colour transformation,
$(U-B)_{STD} = 1.065(U-B)_{OBS} - 0.03$, was found from standard stars with
$(U-B) = -0.17~\rm{to}~+1.12$ and applied to the data. Equatorial coordinates
were obtained as above using IRAF and the USNO-b catalogue.

Photometry was made with SExtractor
(Bertin \& Arnouts 1996), first measuring the deeper $B$ image with the
single image mode and second finding corresponding objects in the $U$ image
by running the double image mode. 

Abell 2465 is included in the DR9 of the SDSS (Ahn \etal 2012). Whilst its
$g', r', i',$ and $z'$ data cover brighter cluster
members, the $u'$ measurements have errors about twice 
those of our $U$ measurements. Consequently, the
current $UB$ data and the SDSS $g', r', i'$ are used.

\begin{figure*}
\centering
\subfloat{\includegraphics[width=0.45\textwidth]{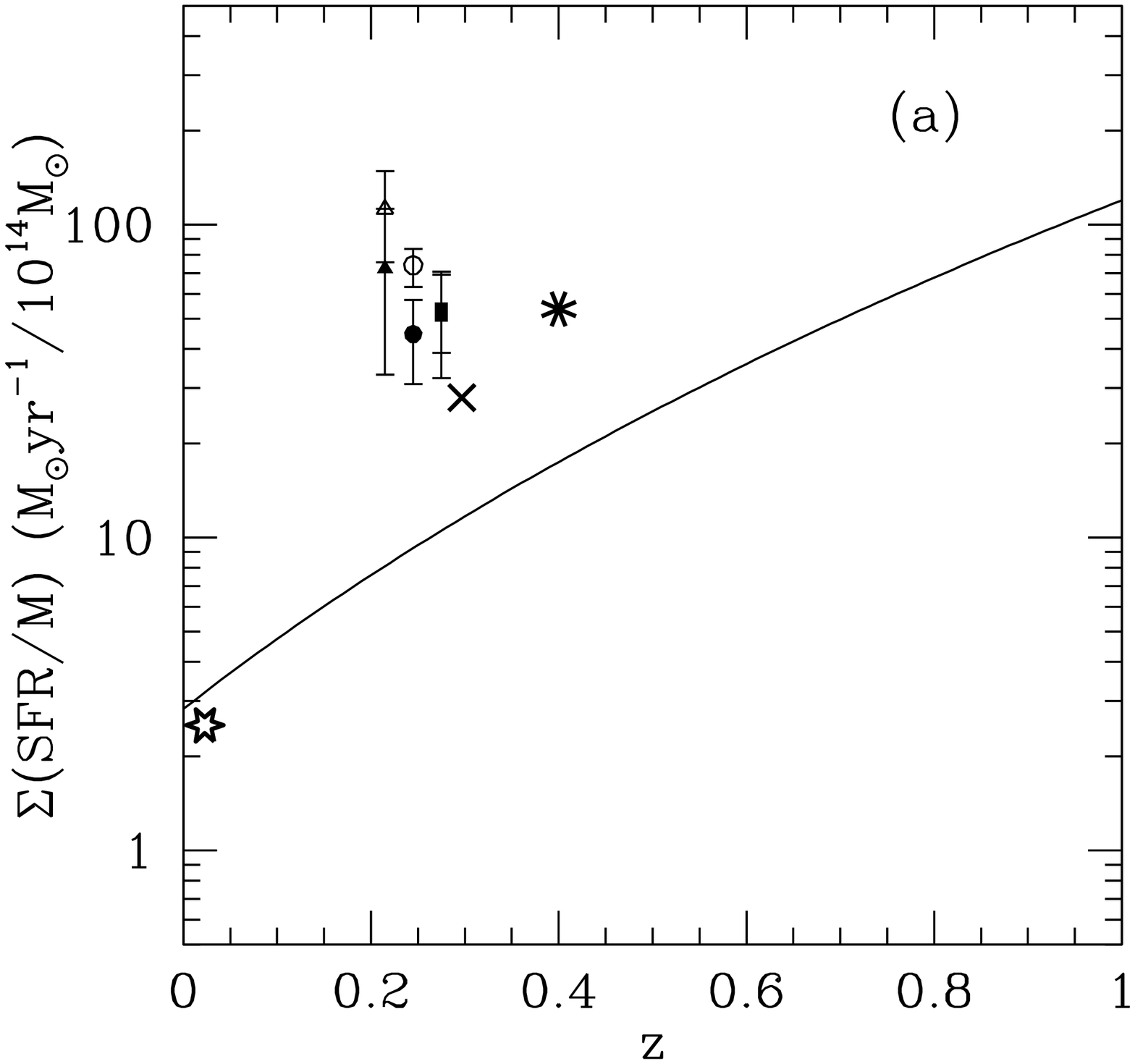}}
\subfloat{\includegraphics[width=0.45\textwidth]{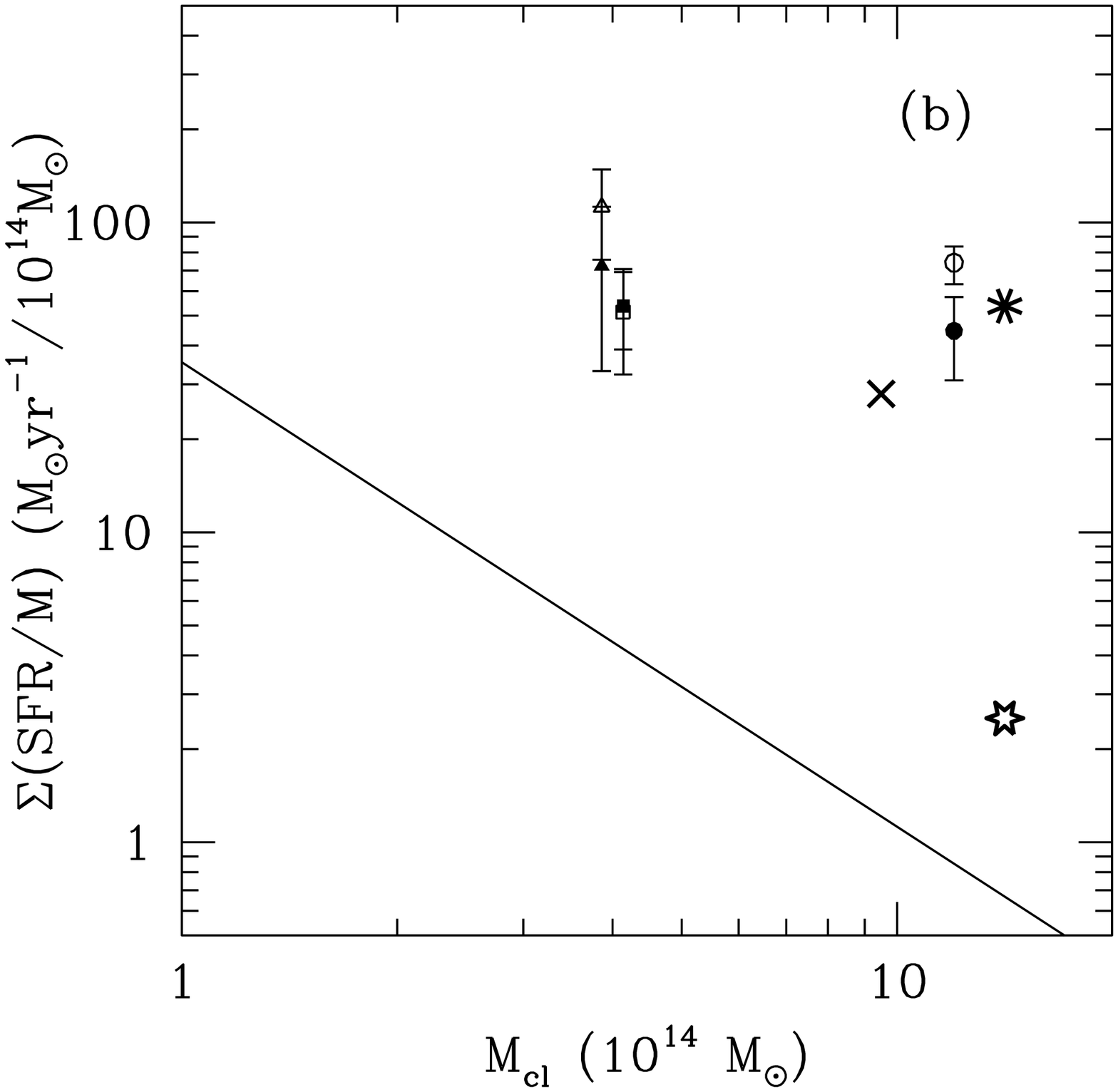}}

\begin{footnotesize}
\caption{Mass normalized SFR inside $R_{200}$
for Abell 2465 compared to the mean relations from Webb \etal (2013)
corrected to $z = 0.245$ and SFR = 0.3 M$_\odot$yr$^{-1}$ as described in the text.
(a) depending on redshift,
$\propto (1 + z)^{5.4}$. 
b) Cluster mass $\propto M_{cl}^{-1.5}$. 
The triangles denote the SW clump, the squares the NE, and
circles the cluster taken as a whole. Open symbols are the IR data and filled symbols are
the H$\alpha$ determinations. The SW and NE points have been displaced 
slightly horizontaly
from their correct positions to enable visibility. The cross shows the position 
of the 'bullet' cluster from Chung \etal (2010), the star Cl0024+1658, and the open star the
Coma cluster.
}
\end{footnotesize}
\label{SFRzM}
\end{figure*}


\subsection{Colour-magnitude diagrams}

\subsubsection{Spectroscopically verified cluster members}
Galaxies with redshifts consistent with membership in Abell 2465 are
in Table 2 of Paper I. We matched the $BU$ measurements with the SDSS
DR9 $g'r'i'$ data.  
The $i',(r'-i')$
diagram in Paper I from CFHT data shows a well defined red 
sequence. Figures~\ref{GmIplot.eps} and \ref{UmBplot.eps} show the
$(i', g'-i')$ and the $(M_B, U-B)$ diagrams for spectroscopically verified 
members and H$\alpha$ sources within $R_{200} = 2$ Mpc. 
In Figure~\ref{GmIplot.eps}, the spectroscopic sample of Paper I, which
covers a larger $15 \times 12$arcmin$^2$ region, is compared to galaxies for 
which H$\alpha$ was measured. The spectroscopic sample was primarily selected using the red
sequence which resulted in most objects being redder than $(g' - i') \ga 1.5$ in
Figure~\ref{GmIplot.eps}a. Possible members in the SW or NE subclusters are assigned
different symbols, but no significant differences in their distributions are seen. The
galaxies with H$\alpha$ in the 2.5$\mathfrak{c}$ sample are in Figure~\ref{GmIplot.eps}b
where a weaker red sequence and more fainter objects bluer than $(g' - i') \la 1.5$ 
are evident.

\subsubsection{The cluster's blue fraction from using the $UB$ data}
The 
$UB$ measurements are useful for comparing the 
blue fractions of Abell 2465 determined for field galaxies and other galaxy clusters. 
The blue fraction determined from the $U$ data is taken to be another indicator of
the star formation activity. For this section, the whole cluster as defined in Sections 3 and 4
will be studied.
The blue fraction, $f_b$ is defined as 
$f_b = (N^{b}_{c+f} - N^b_{f})/(N^{all}_{c+f} - N^{all}_f)$,
where $N^{b}_{c+f}$ and $N^b_{f}$ are respectively the numbers of blue (late-type) 
galaxies 
counted within $0.7R_{200}$ in the cluster and a comparison field;
$N^{all}_{c+f}$ and $N^{all}_f$ are the corresponding numbers of all galaxies in the two
areas. Goto \etal (2003) give an expression for the error, $\delta f_b$, in terms
of these quantities.

The evolution of $f_b$ with $z$ has been extensively studied 
and its rise with $z$ and relation to other cluster properties are well 
known (\eg Margoniner \etal 2001; Goto \etal 2003; De Propris \etal 2004; Tovmassian, 
Plionis \& Andernach 2004; Martinez \etal 2006; Blanton 2006;
Gerke \etal 2007).

The $UB$ measurements of Abell 2465 can be compared with field galaxies.
Using $B$ and $(U-B)$, Cooper \etal (2008) examined field galaxies. For low $z$ they 
divided the red and blue sequences using the restframe division of Willmer \etal (2006):
$$(U-B) = -0.032(M_B+21.52)+0.204$$ 
This uses the $UB$ Vega system (also Baldry \etal 2004).
Galactic extinctions in NED for Abell 2465
and $K$ terms from Fukugita \etal (1995) and Blanton \& Roweis (2007) were applied.
The $B$ measurements reach fainter than the $U$ as seen in Figure~\ref{UmBplot.eps}b
corresponding to a $U$ limit of $M_U \approx -18.0$ mag. Using a
distance modulus of 40.44 mag., this is  
equivalent to $U_{Vega} \approx 22.4$ mag.
The vertical dotted line at $M_B = -19.4$ mag. cuts the data and 
gives a colour independent magnitude limited sample  
from which, the blue galaxy fraction is obtained. For the $UB$ data, the region
centred on the cluster covers 78.9 arcmin$^2$ and two background regions of area 114.7
arcmin$^2$ centered at ($\alpha, \delta)$: (339.73, -5.74) and (339.90, -5.92) were used. 
This yields $f_b = 0.53 \pm 0.02$.

%

A further estimate, of $f_b$ was obtained following
Goto \etal (2003) from the SDSS DR9 $u$ and $r$ data and also
Margoniner \etal (2001). Galaxies were selected within 
$r \le 0.7R_{200}$ (6.09 arcmin). All objects were assumed to have $z = 0.245$, 
were corrected for galactic extinction and $K$-terms. Galaxies
fainter than $M_r$ = -19.44 mag. were excluded and the division between red and blue
sequences was $(u-r) = 2.22$.

Due to the cluster's elongated shape,
rather than a circular annulus, two background
regions of diameter 5.0 arcmin centered at $(\alpha, \delta$): 
(339.6, -5.5) and (340.1, -6.0)
degrees were employed and yields $f_{u-r} = 0.44 \pm 0.04$.   

Noting that systematic errors arise between different galaxy samples and analysis methods,
the estimates of $f_b$ and $f_{u-r}$ for Abell 2465, are compared with the 
average for field galaxies and galaxy clusters at its redshift of $z = 0.245$. 
For the field, Cooper \etal (2008) would indicate $f_b \approx 0.44$.
For galaxy clusters, $f_b \approx 0.3 \pm 0.1$ (Margoniner \etal 2001), while Goto
\etal (2003) find $f_b \approx 0.2$ and $f_{u-r} \approx 0.3$. De Propris \etal (2004)
derive $f_b \approx 0.2$ for spectroscopic cluster members from Two-degree Field Galaxy
Redshift Survey (2dFGRS) data for a
somewhat lower $z < 0.11$ redshift range.
Defining the richness of the cluster (Goto \etal 2003) as the number of galaxies 
corrected for the background within
0.7 Mpc brighter than $M_r = -19.44$ mag., the richness 
number of Abell 2465 is 70 and the correlations for richness of Goto \etal (2003) would
put the expected $f_b$ and $f_{u-r}$ lower than the mean relations.
Although the scatters in these relations are large, Abell 2465 exceeds the mean relations.
Taking $(U - B)$ and $(u-r)$ as indicators of star formation activity
also suggests an increased rate in Abell 2465.

\begin{figure*}
\centering
\subfloat{\includegraphics[width=0.85\textwidth]{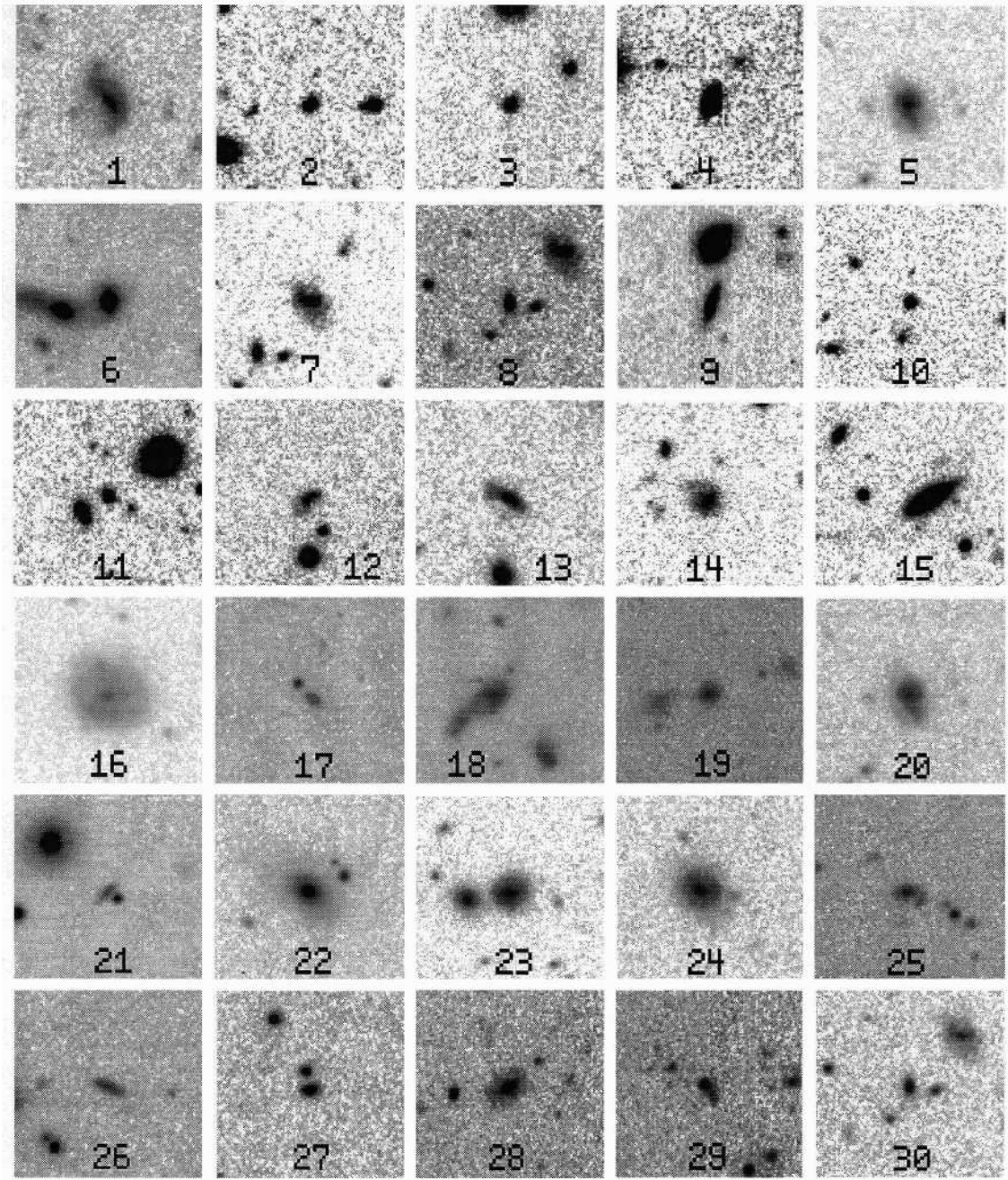}}
\begin{footnotesize}
\caption{Examples of galaxies that have strong H$\alpha$ emission.
Each picture is 20 arcsec or 77 kpc square. North is at the top of each panel and
east is to the left. Images are from the CFHT $i'$ image
described in Paper I. The identities of the galaxies 
(Numbers 1 to 30) in rows from left to right starting with the topmost left are:
J339.8653-5.7788,
J339.8533-5.7717,
J339.9383-5.6769,
J339.9398-5.7923,
J339.9270-5.7072,
J339.7087-5.7743,
J339.8183-5.7337,
J339.8198-5.7352,
J339.7211-5.6597,
J339.9275-5.6508,
J339.8724-5.8025,
J339.8786-5.7371,
J339.8825-5.9239,
J339.8889-5.7428,
J339.8979-5.7721,
J339.9036-5.8588,
J339.9076-5.8043,
J339.9092-5.7782,
J339.9177-5.7150,
J339.9270-5.7073,
J339.9271-5.7474,
J339.9353-5.9162,
J339.9461-5.9085,
J339.9518-5.7792,
J339.9528-5.6837,
J339.8444-5.7669,
J339.8408-5.8449,
J339.8375-5.6874,
J339.8327-5.7795,
J339.8198-5.7352.
}
\end{footnotesize}
\label{empictures}
\end{figure*}

\begin{figure*}
\centering
\subfloat{\includegraphics[width=0.85\textwidth]{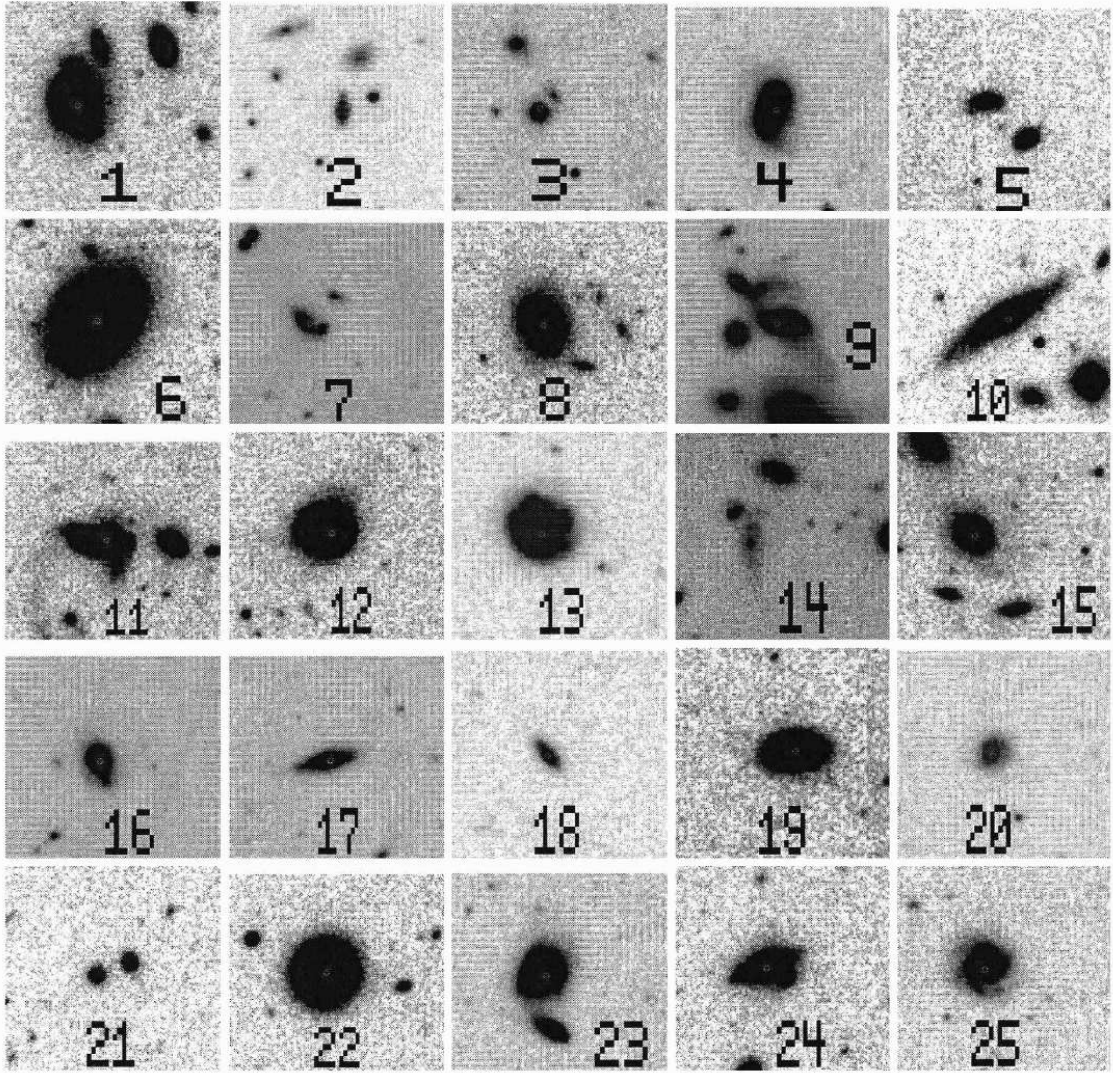}}
\begin{footnotesize}
\caption{Examples of galaxies that were detected by WISE.
Each picture is 20 arcsec or 77 kpc square. North is at the top of each panel and
east is to the left. Images are from the CFHT $i'$ image
described in Paper I. The identities of the galaxies 
(Numbers 1 to 25) in rows from left to right starting with the topmost left are:
J339.7543-5.7144,
J339.7907-5.8257,
J339.8024-5.7229,
J339.8132-5.6724,
J339.8259-5.6249,
J339.8353-5.8227,
J339.8413-5.5778,
J339.8470-5.8618,
J339.8555-5.7813,
J339.8647-5.8962,
J339.8846-5.6863,
J339.8908-5.6856,
J339.9036-5.8589,
J339.9078-5.7287,
J339.9087-5.8407,
J339.9269-5.7073,
J339.9299-5.8067,
J339.9301-5.8939,
J339.9439-5.6529,
J339.9535-5.8746,
J339.9654-5.8267,
J339.9777-5.7796,
J339.9889-5.7557,
J339.9928-5.7687,
J340.0028-5.8228.
}
\end{footnotesize}
\label{WISEpictures}
\end{figure*}



\begin{figure}
\includegraphics[width=0.5\textwidth]{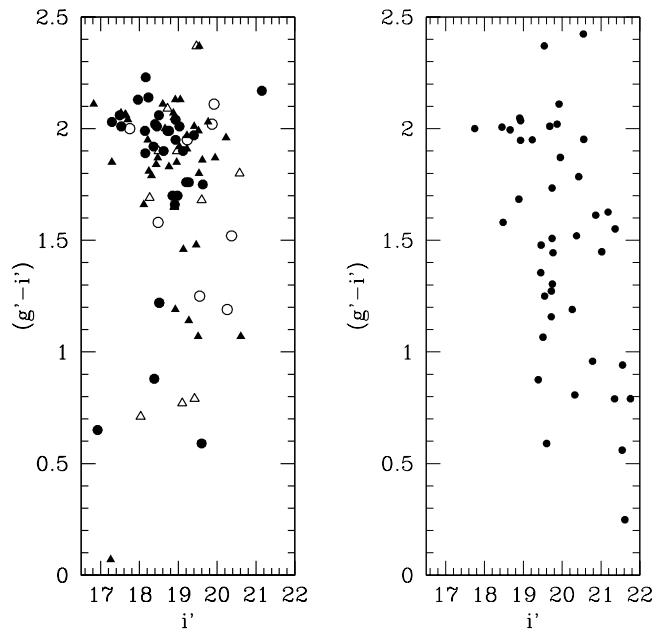}
\begin{footnotesize}
\caption{The $(g'-i')$ and $i'$ colour-magnitude diagram for Abell 2465
from Sloan DR9 data. (left) Spectroscopically identified members. Solid circles
denote Northern clump members with H$\alpha$ emission and filled triangles,
northern members with no detected emission. Open symbols denote the same for southern
clump members. (right) showing members identified with
H$\alpha$ emission detected as described in the text in the central region 
($r < R_{200}$) of the
cluster. Note that the left covers a larger area on the sky. 
}
\end{footnotesize}
\label{GmIplot.eps}
\end{figure}

\begin{figure}
\includegraphics[width=0.5\textwidth]{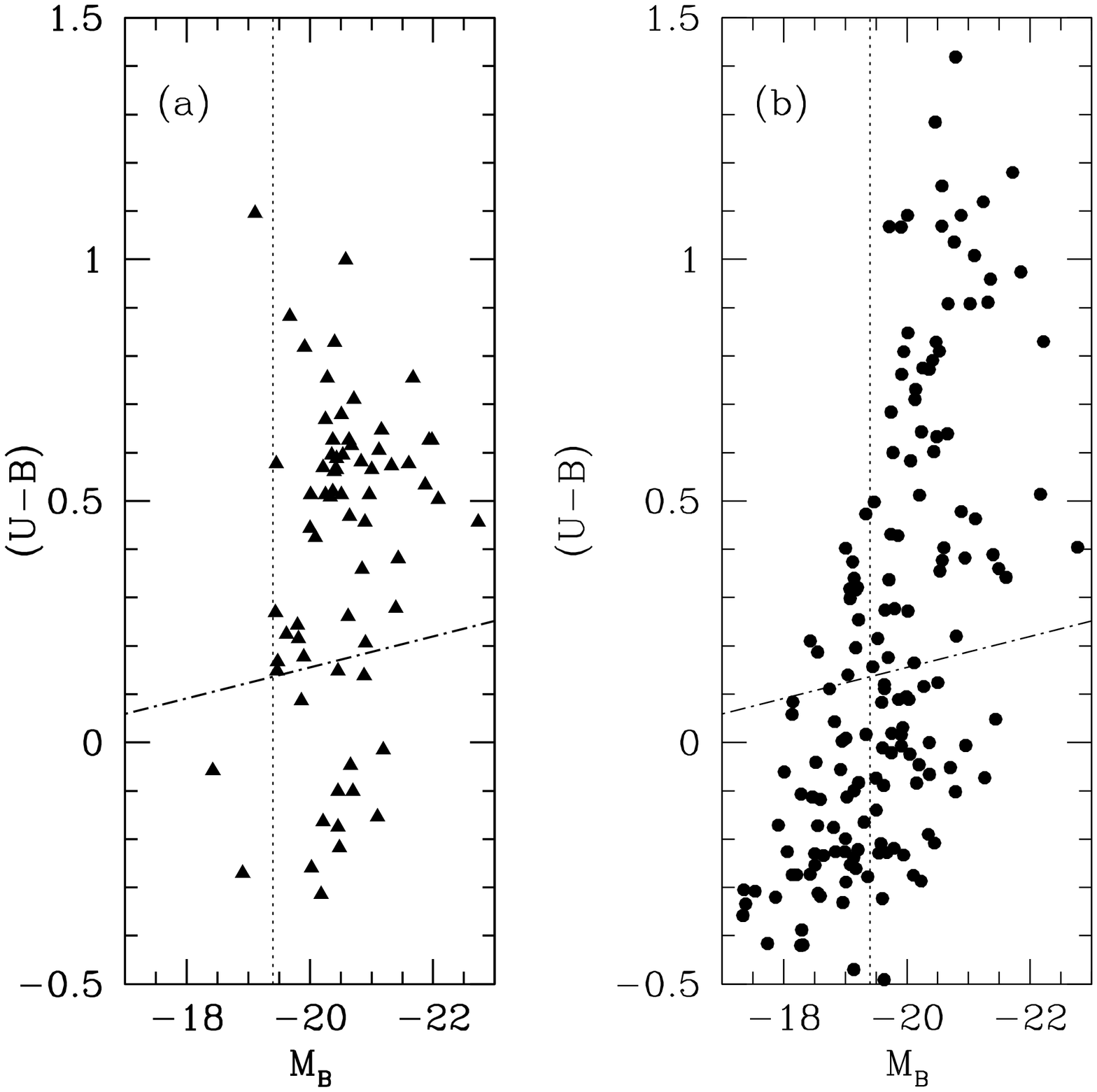}
\begin{footnotesize}
\caption{(a) The $M_B$, $(U-B)$ colour-magnitude diagrams for Abell 2465. (a) 
spectroscopically identified cluster members. (b) The $B, (U-B)$
diagram for all objects that show H$\alpha$ detections inside $r = R_{200}$. 
The vertical dotted line is
the cutoff for the magnitude limited sample. The diagonal dash-dot line is the division 
between the red and blue sequences.  
}
\end{footnotesize}
\label{UmBplot.eps}
\end{figure}

\section{Discussion of possible star formation mechanisms}
Observations and simulations both show that in galaxy clusters, galaxies can rapidly undergo
morphological changes.
Most interactions in galaxy clusters decrease SFR, which
drops with lowering redshift and increasing local density due
to a range of quenching effects. Galaxies are stripped of gas and without renewal of a fresh
supply, the SFR rapidly dies.

So what can trigger 
the apparent SFR excess in some of the double galaxy clusters?
Galaxy evolution processes have been
explored extensively in many studies and reviews (\eg Moore 2004; Boselli \& Gavazzi 2006;
Park \& Hwang 2009; Cattaneo \etal 2011). These processes 
are of two basic types: either dynamical, involving collisional or tidal interactions, or
hydrodynamical stripping. Although each
could operate under proper conditions, observations argue that the
hydrodynamical processes are often most important. Nearby dynamical interactions 
could enhance SF while hydrodynamical ones would quench it. 

The operation of ram pressure stripping during mergers is complicated
(Quilis, Moore, \& Bower 2000; Bekki 1999; 2013). Simulations sometimes
indicate that star formation in galaxies, although usually diminished can be enhanced,
depending on the inclinations and directions of their motion
in the cluster although this is probably not a major effect (Mastropietro, 
Burkert, \& Moore 2008). 
Vijayaragharan \& Ricker (2013) simulated
galaxy groups passing through a larger cluster and predict that ram 
pressure 'pre-processing' operating as far as $\sim 3R_{200}$ can
reduce SFR. The interactions and SFR can thus only be before the first passage through pericentre. 

The conditions in major mergers involving 
galaxy cluster collisions differ from single galaxy clusters as high velocities are possible 
when the cluster centres pass each other and
the large-scale collisional velocities are added to the motions of the galaxies in each cluster. 
The evolution of the double density structures of the merging clusters would produce a
time variable potential, so it is of interest to
reconsider dynamical processes that increas SFR. 
Many galaxies in Figures~\ref{empictures} and \ref{WISEpictures} appear to
be interacting or disturbed. Merging is limited to distances, regions,
and times when galaxies pass
below a maximum velocity and mergers of the two different galaxy families are only possible
when the two overlapping clusters have low enough relative velocity.
A simple estimate for head-on mergers
(Binney \& Tremaine 1987) is that velocity, $V$, 
be below $V~\la~V_{max}$, where $V_{max} \approx 1.2\sqrt{<v^2>}$ 
and $<v^2>$ is the galaxy's internal mean square velocity. For a representative
velocity dispersion of $\sigma = 250$ \kms,
$V_{max} \sim 500$ \kms. Parabolic encounters can merge at
larger separations, but rapid merging is limited to $3~\la~r_h$, where $r_h$ is the 
galactic median radius. 

When the clusters pass through one another velocities could reach thousands of \kms; 
in many cases $V \gg V_{max},$ is expected and multiple high speed
passages, or harassment (Moore \etal 1996) should operate. Many of the objects shown in 
Figures~\ref{empictures} and \ref{WISEpictures} seem to have the properties of harassed 
galaxies. Since SF requires that the galaxies are not stripped of their gas supply, this
limits it to the outer regions of the clusters where many galaxies are falling inward for the
first time. The radial distribution of the SF galaxies in Figure~\ref{radial_SFR} is not as peaked
as an NFW profile and would be consistent with their being distributed in the outer
regions of the cluster.  

Park \& Choi (2009) considered the role of impact interactions between galaxies
and the larger scale density distribution in clusters. They indicate that galaxy-galaxy
collisions will only be important within a limiting radius
and that there is only a weak dependence of galaxy properties on the large-scale
density. This seems to be confirmed, 
\eg by Patton \etal (2013) and others who find that enhanced star formation effectively 
stops beyond a projected separation of $\sim 150$~kpc in a given galaxy pair.


\section{Conclusions}
The goal of this investigation was to determine how the merging of the double components of 
the Abell 2465 galaxy cluster affects the star formation. Consequently the H$\alpha$ and IR
emission of cluster galaxies was measured and converted to SFR and as check, the
blue fraction of galaxies was measured. From this we draw the
following conclusions.

(1) Comparing Abell 2465 with other galaxy clusters for which star formation rates have
been measured using the mass normalized SFR, $\Sigma SFR/M_{cl}$, where $M_{cl}$ is the total
mass inside the virial radius, taken here to be \R200.
Using the available relations as a function of redshift $z$ and and $M_{cl}$, the SW and NE
subclusters and the cluster taken as a whole, show an excess of star formation. This places
Abell 2465 in the regions of the diagrams with other merging galaxy clusters and suggests
that the dynamics of the interaction affect the position of a cluster placing them
higher in the $z$ and $M_{cl}$ diagrams. 

(2) Observations in the $U$ and $B$ bands were obtained of Abell 2465. From the magnitude 
limited sample, the blue fraction of galaxies is 0.53 $\pm$ 0.02, compared to a value 
which would be expected for the field galaxies near 0.2 also indicating a higher star
formation rate. Using $(u-r)$ from the SDSS gives similar results.

(3) The distribution of the star formation,H$\alpha$ and IR sources,
is less concentrated to the centres of the clustar than is the mass which tends to follow an
NFW distribution.

(4) The morphologies of the strongest infrared and H$\alpha$ sources often 
show evidence of interactions. From a consideration of the high velocities possible in the passage
of the two clusters and the strong general evidence for gas stripping in clusters, it is considered
likely that harassment of galaxies in the outer regions of the clusters is operating.


Fundamental to interpreting these observations is knowing whether the merger of the subclusters 
of Abell 2465 has passed or still lies in the future. Unlike hotter and closer cluster pairs 
such as the 'Bullet Cluster' and Cl0024+1658 where the intergalactic gas observed in X-rays 
is hot, the projected separation of the SW and NE centres (5.5 Mpc) and the lower X-ray
temperature $\sim$4 keV (Paper I). It is hoped that X-ray and weak lensing studies now underway will
help resolve some of these questions.  

\section*{Acknowledgements}
Some imaging data in this paper were 
based on observations obtained with MegaPrime/MegaCam, a joint project of CFHT 
and CEA/DAPNIA, at the Canada-France-Hawaii Telescope (CFHT) which is operated 
by the National Research Council (NRC) of Canada, the Institut National des 
Science de l'Univers of the Centre National de la Recherche Scientifique (CNRS)
of France, and the University of Hawaii. This work is based in part on data 
products produced at TERAPIX and the Canadian Astronomy Data Centre as part of 
the Canada-France-Hawaii Telescope Legacy Survey, a collaborative project of 
NRC and CNRS. I thank the Canadian TAC for granting the 
time and the QSO team for obtaining the imaging data.
This research has made use of the NASA/IPAC Extragalactic Database (NED) which 
is operated by the Jet Propulsion Laboratory, California Institute of 
Technology, under contract with the National Aeronautics and Space 
Administration. Many thanks to Dr. Dane Owen who helped with the $UB$ imaging at MDM Observatory.


\label{lastpage}

\end{document}